\documentclass[final,3p,times,comma,round,numbers,twocolumn,sort&compress]{elsarticle}

\usepackage{xr}
\usepackage{hyperref}
\usepackage{amsmath}
\usepackage{amssymb}
\usepackage{amsfonts}
\usepackage{graphicx}
\usepackage{dcolumn}
\usepackage{bm}
\usepackage{times}
\usepackage{xcolor}
\usepackage{longtable}
\usepackage{chapterbib}
\usepackage[normalem]{ulem}
\usepackage{tabularx}
\usepackage{chapterbib}

\newcommand{\squeezeup}{\vspace{-2.5mm}}
\newcommand{\change}[1]{{\color{black}#1}}
\newcommand{\changen}[1]{{\color{black}#1}}
\journal{Physical Review B}

\begin{document}

\title{A structural study of hcp and liquid iron under shock compression up to 275 GPa}

\author[a]{Saransh Singh\corref{cor1}}
\ead{saransh1@llnl.gov}
\cortext[cor1]{Corresponding author}
\author[a]{Richard Briggs}
\author[a]{Martin G. Gorman}
\author[a]{Lorin X. Benedict}
\author[a]{Christine J. Wu}
\author[a]{Sebastien Hamel}
\author[a]{Amy L. Coleman}
\author[a]{Federica Coppari}
\author[b]{Amalia Fernandez-Pa\~{n}ella}
\author[a]{Christopher McGuire}
\author[c]{Melissa Sims}
\author[c]{\\June K. Wicks}
\author[a]{Jon H. Eggert}
\author[a]{Dayne E. Fratanduono}
\author[a]{Raymond F. Smith}

\affiliation[a]{Lawrence Livermore National Laboratory, Livermore, CA 94550, USA}
\affiliation[b]{Gordon and Betty Moore Foundation, Palo Alto, CA 94304, USA}
\affiliation[c]{Department of Earth and Planetary Sciences, Johns Hopkins University, Baltimore, Maryland 21218, USA}

\date{\today}

\begin{abstract}
	\noindent
	We combine nanosecond laser shock compression with \emph{in-situ} picosecond X-ray diffraction to provide structural data on iron up to 275 GPa. We constrain the extent of hcp-liquid coexistence, the onset of total melt, and the structure within the liquid phase. Our results indicate that iron, under shock compression, melts completely by 258(8) GPa. A coordination number analysis indicates that iron is a simple liquid at these pressure-temperature conditions. We also perform texture analysis between the ambient body-centered-cubic (bcc) $\alpha$, and the hexagonal-closed-packed (hcp) high-pressure $\epsilon-$phase. We rule out the Rong-Dunlop orientation relationship (OR) between the $\alpha$ and $\epsilon-$phases. However, we cannot distinguish between three other closely related ORs: Burger's, Mao-Bassett-Takahashi, and Potter's OR. The solid-liquid coexistence region is constrained from a melt onset pressure of 225(3) GPa from previously published sound speed measurements and full melt (246.5(1.8)-258(8) GPa) from X-ray diffraction measurements, with an associated \change{maximum} latent heat of melting of \change{623} J/g. This value is lower than recently reported theoretical estimates and suggests that the contribution to the earth's geodynamo energy budget from heat release due to freezing of the inner core is smaller than previously thought. Melt pressures for these nanosecond shock experiments are consistent with gas gun shock experiments that last for microseconds, indicating that the melt transition occurs rapidly.
\end{abstract}

\maketitle
\squeezeup
\squeezeup
\section{Introduction}
Iron is a cosmochemically abundant element that plays a significant role in terrestrial planetary interiors as the dominant core constituent. The ultra-high pressure properties of iron are essential for interpreting the dynamics and interior structure of the earth and rocky exoplanets \cite{anzellini2013,Smith2018,Kraus2022}. Within the earth, it is estimated that the solid inner core is comprised of Fe alloyed with $\sim$5-10\% of impurities (e.g., Si, S, O, C, H, and Ni) by weight \cite{Hirose2021}. 
Surrounding the solid inner core is the Fe-rich outer liquid core, which is estimated to have $\sim$8-16$\%$ impurity content by weight \cite{Hirose2021}.~According to the standard model, convection in the outer core is driven by processes associated with solidification and growth of the inner core. One source is the buoyancy generated by the exclusion of incompatible light elements from the solid. Another is latent heat release from re-crystallization.~Planetary magnetic fields arising from convection within the outer liquid Fe-rich cores play an important role in the atmospheric evolution and surface environment of planets \cite{Foley2016}.~There is a strong need to constrain material properties of Fe close to melting at the extreme pressures found within planetary interiors to understand these processes better. 

High-pressure static compression studies on Fe have revealed a phase transformation from the ambient body-centered-cubic (bcc) $\alpha$-phase to a hexagonal-closed packed (hcp) $\epsilon$-phase at 15.3 GPa with an associated 5\% volume collapse \cite{barge1990, boehler1990, taylor1991}. A large (shear-stress dependent) hysteresis of the transition is observed with a midpoint pressure of 12.9 GPa \cite{barge1990, boehler1990, taylor1991}. The melting of Fe has also been reported as a function of pressure by static compression techniques \cite{anzellini2013,Sinmyo2019,Kuwayama2020,boehler1993,boehler1990}, and theoretical calculations \cite{belonoshko2000,alfe2009,swift2020}.~Under near instantaneous uniaxial shock compression, the $\alpha$$\rightarrow$$\epsilon$ phase transformation in polycrystalline Fe samples has been observed, after a period of stress relaxation, to initiate at 12.9 GPa \cite{bancroft1956,Jensen2009,barker1974,barker1975}. At higher pressure, Nguyen and Holmes \cite{nguyen2004} used changes in sound speed within mm-thick samples shock compressed over microseconds to infer the onset of melt at 225(3) GPa, and its completion by 260(3) GPa. \change{However, due to the considerable spread in that data, an independent measurement of the Hugoniot intersection with the melt curve is needed. This was \changen{undertaken} by Turneaure et al. \cite{Turneaure2020}}.~Under nanosecond laser shock-compression of 15-$\mu$m thick samples, Turneaure et al. reported melt onset between 241.5(3)-242.4(2.3) GPa through \emph{in situ} X-ray diffraction (XRD) measurements. \change{However, melt completion was constrained only over a relatively wide pressure range, between 246.5(1.8) - 273(2) GPa.}~A further constraint on the high-pressure Fe melt line was reported in a recent laser-driven shock-ramp XRD experiments by Kraus et al. \cite{Kraus2022}. \changen{Finally, recent sound speed measurements by Zhang et al. \cite{zhang2023} shows a smooth and gradual change in sound speed with increasing pressure. Contrary to earlier measurements by Nguyen and Holmes \cite{nguyen2004}, the authors report no sharp drop in the sound speed up to 230.8(1) GPa, indicating no first-order phase transition around the currently accepted melt onset pressure of 225(3) GPa.}

In the experiments reported here, we employ nanosecond laser shock compression combined with \emph{in situ} picosecond XRD and velocimetry techniques to \change{study the structure of iron up to 275 GPa. \changen{This experimental configuration is similar to that used in Ref.~\cite{Turneaure2020}}. We use a forward model for texture analysis at lower pressures to rule out the Rong-Dunlop orientation relationship during the $\alpha\rightarrow\epsilon$ phase transition. At higher pressures, we show that Fe is fully melted along the Hugoniot at 258(8) GPa.~We report on the first liquid structural and density measurements of Fe under the earth's core conditions. Our data provides an independent measurement of the melt completion pressure of iron on the Hugoniot and is largely consistent with previous determination of this pressure using sound speed measurements. Our melting pressure determination, combined with previous sound-speed data \cite{nguyen2004} and a semi-empirical equation of state model \cite{Wu2023}, provides a constraint on the latent heat of fusion of iron at extreme conditions present in the earth's interior.}

\begin{figure}[t!]
	\centering
	\includegraphics[width=0.95\columnwidth]{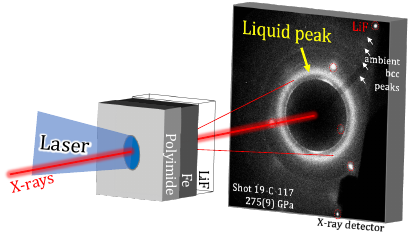}
	\caption{\label{fig:setup}
		\textbf{$\vert$ Experimental setup.} The target design consists of a polyimide ablator, a 21-$\mu$m-thick Fe foil, and a LiF window for velocimetry measurements \cite{dolan2006}. Raw diffraction data from laser shock compressed Fe at an estimated pressure of 275(9) GPa.
	}
	\squeezeup
	\squeezeup
\end{figure}

\section{Experimental Method}
\squeezeup
We conducted laser shock compression experiments at the Dynamic Compression Sector (DCS) beamline of the Advanced Photon Source (APS), located at the Argonne National Laboratory \cite{wang2019}. 
Targets consisted of a polyimide ablator ($\sim$35-50 $\mu$m thickness) glued to high purity 21 $\mu$m-thick Fe foil (see Fig. \ref{fig:setup}). An [100] orientated LiF window was glued to the Fe sample to facilitate measurements of the Fe/LiF particle velocity.
A 1-3 $\mu$m-thick epoxy held these three layers together. The LiF window is coated with 0.1 $\mu$m-thick Ti to enhance reflectivity for velocimetry measurements (VISAR) \cite{dolan2006}. The full density 99.99\% pure Fe foils were supplied by Goodfellow, USA (initial density = 7.874 g/cm$^3$). 

A 5 or 10 ns approximately flat-top laser pulse at 351 nm, with energies between 15 and 80 J, was focused within a 580 $\mu$m diameter focal spot on the front surface of the polyimide layer.
~This setup uniaxially shock-compressed the target assembly and the generated longitudinal stress states in Fe between 25 and 275 GPa.~A point-VISAR Doppler velocity interferometer was used to measure the time history of the Fe/LiF interface, $u_p$(t) \cite{dolan2006}, and through standard impedance-matching techniques, this allowed a determination of shock pressure in the iron sample. Simultaneously, an X-ray pulse ($\sim$80 ps) was timed to probe the compressed sample during shock transit within the Fe, which produced an X-ray diffraction pattern recorded in a transmission geometry (Fig.~\ref{fig:setup}), with contributions from the compressed Fe (behind the shock front) and uncompressed Fe (ahead of the shock front). The X-ray flux had a peak at an energy of $\sim$23.56 keV with a pink beam profile. The experimental geometry has also been described in Refs.~\cite{wang2019,briggs2019,Coleman2022,Sims2022}.
\squeezeup
\subsection{X-ray diffraction data processing}
\begin{figure}[t!]
	\begin{center}
		\includegraphics[width=\columnwidth]{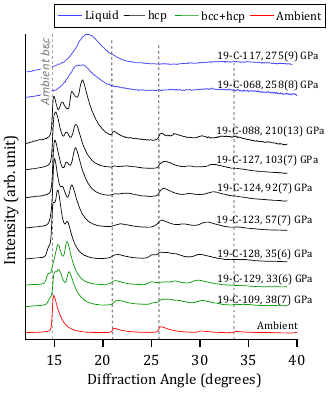}
		\caption{\label{fig:XRD}\textbf{X-ray diffraction} Azimuthally averaged diffraction data as a function of increasing shock pressure (right axis). The dashed vertical lines are diffraction angles for the ambient bcc Fe. The traces are colored coded to depict crystal structure: bcc+hcp (green), hcp (black), and liquid (blue). \change{See  Figs. \ref{fig:Lineout_1}-\ref{fig:Lineout_3} for LeBail fits to averaged diffraction lineouts used to determine crystal structure}.
		}
	\end{center}
	\squeezeup
	\squeezeup
	\squeezeup
\end{figure}

CeO$_{2}$, and powdered Si calibrants were used to determine the sample to detector distance, beam center, tilt, and rotation.~XRD images were azimuthally averaged and analyzed using HEXRD python package \cite{HEXRDgithub}. First, we used powder diffraction rings from known standards (CeO$_2$ and Si) for detector calibration. The pink beam profile function detailed in Ref. \cite{vondreele2021} was used for this step. Next, we performed several intensity corrections to enable accurate density determinations from the diffraction profiles. The intensity corrections included (i) subtracting the dark counts, (ii) subtracting the ambient Fe pattern for the case of liquid XRD analysis and subtracting LiF single-crystal Laue peaks for all experiments, (iii) correcting for the solid angle subtended by each pixel on the detector (iv) accounting for the polarization factor (for liquid diffraction profiles only) and (v) correcting for the attenuation due to varying path length of the diffracted X-rays. Figure~\ref{fig:XRD} shows representative intensity corrected integrated diffraction profiles for shock-compressed Fe (see also Figs.~\ref{fig:Lineout_1}-\ref{fig:Lineout_3}). 
We increased the incident drive laser energy to achieve stresses on the Hugoniot between 25 and 275 GPa and observed a range of azimuthally averaged x-ray diffraction lineouts, indicating different Fe structures.~Over the stress range studied, Fe exhibits an hcp structure from 25-210 GPa. For the two highest pressure shots (258 and 275 GPa), diffuse scattering around Q = 3.8 $\AA^{-1}$ is present in these profiles with no evidence of compressed hcp; this is characteristic of the complete melting [Q = 4$\pi$sin($\theta$)/$\lambda$, where $\lambda$ is the X-ray wavelength, and $\theta$ is the Bragg scattering angle].

\subsection{Stress determination}
\begin{figure}[h!]
	\begin{center}
		\includegraphics[width=\columnwidth]{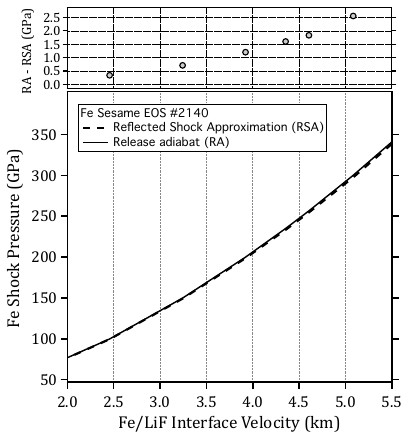}
		\caption{\textbf{Pressure determination using Fe-LiF impedance matching.} Using Sesame EOS table \changen{\#2150} for Fe for a range of assumed LiF u$_p$ values, the main plot shows the pressure in Fe calculated through impedance-matching assuming (i) the reflected shock approximation (RSA) and (ii) the release adiabat (RA) approach. In these calculations, the RA path does not include a strength model. While the RA approach is physically correct, the RSA is often used in impedance matching because it simplifies the analysis, as one needs only the Hugoniot data to perform the analysis. In contrast, the RA approach relies on a theoretical model \cite{celliers2005,kerley2013}. The upper plot represents an underestimation of the calculated Fe pressure from the RSA approach as a function of increasing pressure. We have accounted for this offset in our pressure estimates.
		}
		\label{fig:Impedance}
	\end{center}
	\squeezeup
	\squeezeup
\end{figure}

\begin{figure*}[h!]
	\begin{center}
		\includegraphics[width=0.85\textwidth]{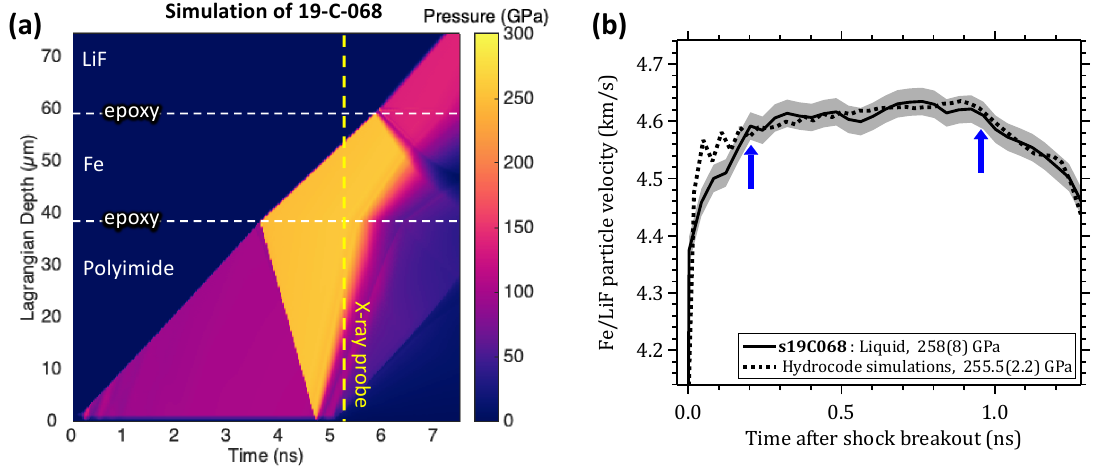}
		\caption{{\color{black}\textbf{$\vert$ Hyades simulation of shot 19-C-068.} \textbf{a.} Calculated map of pressure distribution throughout the target package, where the horizontal dashed lines indicate material boundaries in Lagrangian coordinates. For this shot, the 80 ps X-ray pulse probed the sample 0.7 ns before shock breakout into the LiF window. The resultant X-ray diffraction pattern is volume integrated and, therefore, includes a scattering contribution from the ambient material ahead of the shock front and the shocked material (behind the shock front). In this simulation, Fe EOS table \#2150 from Ref. \cite{kerley1993} (see also Fig. \ref{fig:2150}), and \#7271 for LiF \cite{Davis2016} were used. \textbf{b.} Agreement between measured (solid) with calculated $u_{p,LiF}$($t$) (dashed). We obtain a value of 258(8) GPa using the impedance matching approach for determining pressure. We consider the velocity distribution over the period between the two blue arrows for this estimate. These peak velocity values are most closely associated with the location of the peak of the diffraction data, which is used to calculate density. During the X-ray probe period, the hydrocode simulations calculate a value of 255.5(2.2) GPa, consistent with the impedance matching estimates.}
		}
		\label{fig:Hyades}
	\end{center}
	\squeezeup
	\squeezeup
\end{figure*}

We employed standard impedance matching techniques to determine the stress in the Fe layer from the measurement of the Fe/LiF particle velocity, and using $U_S$-$u_{p}$ fits (shock velocity-particle velocity, both in $km/s$)  to \change{previously measured} shock compression data for LiF \cite{Davis2016,Rigg2014,hawreliak2023} (Eqn. \ref{eq:LiF_Us_up}) and Fe \cite{Walsh1957,Altshuler1958,Altshuler1962,Altshuler1968,Altshuler1974,Altshuler1977,Altshuler1981,Altshuler1996,Altshuler1960,Mcqueen1960,Mcqueen1970,Skidmore1962,Krupnikov1963,Balchan1966,Ragan1984,Hixson1991,Brown2000,Trunin1972,Trunin1992,Trunin1993,Trunin1994,Trunin2001} (Eqn. \ref{eq:Fe_Us_up}). We calculated the Fe/LiF particle velocity after accounting for the refractive index of LiF under shock compression \cite{Rigg2014}.

\medskip
\begin{equation}
	U_{S, LiF} = 5.215 (\pm 0.048) + 1.351 (\pm0.025) \times u_{p, LiF}
	\label{eq:LiF_Us_up}
\end{equation}
\begin{equation}
	U_{S, Fe} = a + b \times u_{p, Fe} - c \times u_{p, Fe}^2 + d \times u_{p, Fe}^3
	\label{eq:Fe_Us_up},
\end{equation}

\medskip
\medskip
\noindent
where a = 3.4188 ($\pm$0.0539), b = 2.1663 ($\pm$0.0823), c = 0.1992 ($\pm$0.0369), and d = 0.021219 ($\pm$0.00497). The uncertainties from impedance matching have contributions from the scatter of EOS data for both Fe \cite{Walsh1957,Altshuler1958,Altshuler1962,Altshuler1968,Altshuler1974,Altshuler1977,Altshuler1981,Altshuler1996,Altshuler1960,Mcqueen1960,Mcqueen1970,Skidmore1962,Krupnikov1963,Balchan1966,Ragan1984,Hixson1991,Brown2000,Trunin1972,Trunin1992,Trunin1993,Trunin1994,Trunin2001} and LiF \cite{Davis2016,Rigg2014,hawreliak2023} and the steadiness of the shock wave as determined in the velocity histories (Fig.~\ref{fig:VISAR}a,c and e). Reported uncertainty in the Hugoniots for Fe and LiF to place a \emph{minimum} uncertainty of $\pm$0.4 GPa for Fe shock stress determination through impedance matching. This value was calculated for $P_{Fe}$=250 GPa, based on fits to Hugoniot data with 1-$\sigma$ confidence bands. Confidence bands rely only on the fit coefficients and the estimated uncertainties in the coefficients. This value represents a systematic uncertainty and is combined with estimated random uncertainties (described below) to give the total pressure uncertainty values reported in Table \ref{table:Summ_table}. While the use of reflected Hugoniots is a standard approximation for the pressure determination through impedance matching \cite{kerley2013,celliers2005}, we note that at high pressure, the use of a release adiabat is needed (Fig.~\ref{fig:Impedance}). In our analysis, we initially conducted impedance matching using the reflected shock approximations and then made an additional correction based on the trend in Fig. \ref{fig:Impedance}, upper.

The measured Fe/LiF particle velocity profiles, shown in Figure~\ref{fig:VISAR}, are characterized by an initial shock followed by a time-dependent distribution of velocity states. The stress in the sample (as reported in Table \ref{table:Summ_table}) is calculated through impedance matching {\color{black}while considering} the velocity distribution after the initial shock {\color{black}(see Fig. \ref{fig:Hyades}b)}. The calculated {\color{black} random} uncertainty in stress is a contribution of the following: (i) the standard distribution of velocity states above the initial shock, and (ii) the accuracy with which fringe shifts can be measured in the point-VISAR system, taken here as 0.024 km/s (2\% of a fringe shift \cite{dolan2006}). Other contributors to stress uncertainty that is not explicitly treated relate to uncertainties in the refractive index of LiF \cite{Rigg2014}, uncertainties in the timing of the X-ray probe with respect to the VISAR, and uncertainties in the measurements of sample thickness. In addition, the point VISAR system integrates spatially and, therefore does not provide any information on the distribution of stress states that may arise due to non-planarities within the drive (as in Ref. \cite{Coleman2022}. Moreover, in the case of a detected Fe/LiF velocity profile, which is characterized by an initial shock followed by an increase in velocity, there is a progressive increase in shock strength throughout the sample as late-time characteristics catch up and reinforce, to some extent, the shock front during transit. To account for these additional uncertainties, we increase our total pressure uncertainties by an additional $\pm$ 5 GPa. We note that our pressure uncertainties are similar to Ref. \cite{Coleman2022} but are significantly higher than reported in Refs. \cite{Turneaure2020, Sharma2019}. The high-pressure constraint for solid-liquid coexistence in Fe is shot 19-C-068, which is the lowest pressure observation of liquid only. The calculated pressure for this shot based on impedance matching as described above is 258(8) GPa.

As an additional check on the pressure and pressure distribution within the sample during the X-ray probe time, the pressure history in Fe is also determined by simulating the experimental conditions with a 1D hydrocode, HYADES \cite{Larsen1994}, which calculates the hydrodynamic flow of pressure waves through the target assembly in time ($t$) and Lagrangian space ($h$) (Fig. \ref{fig:Hyades}a). The inputs to the hydrocode are the thicknesses of each of the constituent layers of the target, including the measured epoxy layer thicknesses ($\sim$1-3-$\mu$m), an equation-of-state (EOS) description of each of the materials within the target, and laser intensity as a function of time, $I_{Laser}$($t$). Based on experience over hundreds of shots, we find that pressure (GPa) in the polyimide ablator scales approximately as 4.65$\times$$I_{Laser}$$^{0.8}$, with $I_{Laser}$($t$) (PW/m$^2$) calculated from measurements of laser power divided by an estimate of the laser spot size (Fig.~\ref{fig:DCS}). ~We ran a series of forward calculations with iterative adjustments of $I_{Laser}$($t$) (few \% level) until convergence was reached between the calculated $u_{p,LiF}$($t$) and the measured $u_{p,LiF}$($t$) curves (Fig. \ref{fig:Hyades}b). Once achieved, we used the calculated $P$($h$,$t$), 
at the X-ray probe time to estimate the volume-integrated pressure and pressure distribution in the Fe sample. Our pressure determination method explicitly accounts for any temporal non-steadiness in the compression wave. The results of the hydrocode simulations in Fig.~\ref{fig:Hyades} are for s19-C-068, which in our experiments represents the first liquid-only shot (the upper-pressure constraint of the solid-liquid coexistence). The determined pressure in the Fe sample from the hydrocode simulations is 255.0(2.2) GPa, which is in agreement with the impedance matching approach (258(8) GPa) described above.

\section{Results}
\changen{We present the results of our shock compression study in this section. Instead of organizing the results with increasing pressure, we have organized the section with what we considered the most impactful. First, we present the highest pressure liquid structure analysis, followed by the texture analysis of the $\alpha$ $\rightarrow$ $\epsilon$ phase transition with implications to plasticity at extreme compression rates. Finally, we present the full Hugoniot measurements and their implications for the iron phase diagram.}
\squeezeup
\subsection{Liquid structure measurements}
The high flux of the X-ray source at DCS coupled with angular coverage up to almost 8 $\AA^{-1}$ permits quantitative analysis of liquid density recorded using \textit{ps} X-ray diffraction. Due to the non-monochromatic ``pink" X-ray beam, the raw liquid diffraction intensities are artificially shifted to higher Q \cite{Bratos2014}. The corrections outlined in Ref. \cite{singh2022} were applied to account for this artificial shift and derive quantitative densities. We have used this method to study other \change{elemental} metallic systems such as Ag, Sn, and Cu. The derived densities agree well with the Hugoniot pressure-density relationship \cite{Coleman2022,singh2022,Sims2022}. 
\begin{figure}[t!]
	\centering
	\includegraphics[width=0.95\columnwidth]{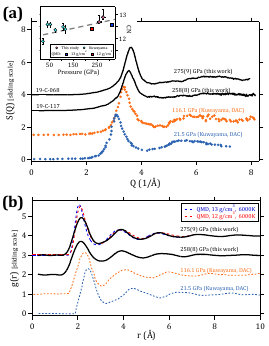}
	\caption{\label{fig:Liquid}
		\textbf{$\vert$ Liquid structure measurements.} \textbf{a.} Liquid structure factor (solid lines) and \textbf{b.} corresponding radial distribution function (solid lines) for dynamically compressed Fe at 258(8) GPa (s19-C-068) and 275(9) GPa (s19-C-117). Also shown is liquid XRD data taken at lower pressure under laser-heated static compression (markers and dashed lines) \cite{Kuwayama2020}. \change{ The red and blue dashed lines are derived from QMD simulations in Ref.~\cite{Wu2023}. Inset in \textbf{a.} presents the coordination number (CN) as a function of pressure. The cyan circles were derived from the published density and radial distribution function from Ref.~\cite{Kuwayama2020}. The pink circles are from this study. The cyan and red square symbols are CNs derived from QMD simulations for two different densities.}
	}
	\squeezeup
	\squeezeup
\end{figure}

Figures~\ref{fig:Liquid}a and b presents the structure factor, $S(Q)$, and the corresponding radial distribution functions, $g(r)$ of the two shots with complete melting (solid black lines).~We show the lowest and highest pressure liquid structure data presented in Ref.~\cite{Kuwayama2020} (markers and dashed lines).~That data was collected using a laser-heated diamond anvil cell apparatus up to a maximum pressure of 116 GPa, providing the density of liquid iron along several isotherms that lie close to the melting temperature \cite{Kuwayama2020}. The $S(Q)$ and $g(r)$ follow the expected behavior of shifting to higher $Q$ and lower $r$.

\changen{We also compare our experimentally determined radial distribution function to quantum molecular dynamic (QMD) simulations, representing the best simulation technique to study structures at these extreme conditions. A rigorous comparison of theory and experiments at these conditions points provides essential information for modelers to fine-tune their techniques.}\change{Based on the QMD simulations reported in Ref. \cite{Wu2023}, the radial distribution functions for two different densities are shown (blue and red dashed lines in Fig.~\ref{fig:Liquid}b). The simulations were performed at the same temperature of $6000$ K. While there is good agreement between the peak position of the QMD and the experimentally determined liquid radial distribution functions, there is disagreement between the shape of the first peak. This indicates a difference in the distribution of atoms in the first coordination shell.} 

We determined the coordination number (CN) by integrating the area under the first peak in the radial distribution function. We followed two different integration schemes: (i) twice the area up to the first peak and (ii) the area up to the first minimum when plotting 4$\pi$nr$^{2}$g(r) against r (n is the number density). We find that the coordination numbers at 258(8) and 275(9) GPa are $\sim$11.7 and 12.2, respectively, using the first method, and 12.7 and 12.9 using the second method \cite{waseda1980}. Either method shows simple liquid behavior. \change{We show the change in coordination number (CN) as a function of pressure using the second method in Fig.~\ref{fig:Liquid}a, inset. The cyan circles denote the CN from the density and radial distribution function reported in Ref. \cite{Kuwayama2020}, \change{while the blue and red square symbols are CNs derived from QMD simulations at $12$ and $13$ g/cm$^3$ and $6000$ K \cite{Wu2023}. Even though there is disagreement in the shape of the first peak of the radial distribution function between the QMD simulation and the experiments, the CNs given by the area of the first peak have comparable values. 
	}~The CN shows a general upward trend with increasing pressure. }

\subsection{\changen{$\alpha\rightarrow\epsilon$ solid-solid phase transition}}
\medskip

\newcolumntype{b}{>{\hsize=0.5\hsize}X}
\newcolumntype{s}{>{\hsize=0.25\hsize}X}
\begin{table*}
	\caption{\textbf{$\vert$ Commonly observed orientation relationships between bcc and hcp phases}\label{Table:OR_summ}}
	\begin{tabularx}{\textwidth}{|s|b|s|}
		\hline
		\centering
		\textbf{Authors} & \textbf{Details} & $\pmb{(\textrm{hkl})_{\alpha}}~~~~ \vert\vert~~~~ \pmb{(\textrm{hkl})_{\epsilon}} \newline \pmb{[\textrm{uvw}]_{\ \alpha}}~~ \vert\vert~~~~ \pmb{[\textrm{uvw}]_{ \epsilon}}$ \\
		\hline
		Burger's \cite{burgers1934}\newline (1934) &  Experimental X-ray diffraction study on the orientation relationship between the hcp $\alpha-$ and bcc $\beta-$phase in Zr. & $(110)_{\alpha}~~ \vert\vert ~~(0001)_{\epsilon}$ \newline $[1\bar{1}1]_{\alpha}~~~ \vert\vert ~~ [11\bar{2}0]_{\epsilon}$ \\
		\hline
		Mao-Bassett-Takahashi \cite{Mao1967} \newline (1967) & Experimental X-ray diffraction study on the orientation relationship between the ambient $\alpha-$phase and high-pressure $\epsilon-$phase in Fe & $(110)_{\alpha}~~ \vert\vert ~~(0001)_{\epsilon}$ \newline 
		$[001]_{\alpha}~~~ \vert\vert ~~[2\bar{1}\bar{1}0]_{\epsilon}$ \\
		\hline
		Potter \cite{potter1973} \newline (1973) &  Experimental study on orientation relationship between bcc $\alpha-$phase in Vanadium-Nitrogen system and hcp V$_{3}$N precipitate using transmission electron microscopy & $(110)_{\alpha}~~ \vert\vert ~~(1\bar{1}01)_{\epsilon}$ \newline $[1\bar{1}1]_{\alpha}~~~ \vert\vert ~~[11\bar{2}0]_{\epsilon}$ \\
		\hline
		Rong-Dunlop \cite{rong1984} \newline (1984) &  Experimental study on orientation relationship between bcc $\alpha-$ferrite and hcp M$_{2}$C (M = Cr, Mo, Fe) precipitates in ASP23 high strength steel using a transmission electron microscope & $(021)_{\alpha} ~~\vert\vert~~ (0001)_{\epsilon}$ \newline $[100]_{\alpha}~~ ~\vert\vert ~~[11\bar{2}0]_{\epsilon}$ \\
		\hline 
	\end{tabularx}
\end{table*}
Iron undergoes bcc $\alpha$ to hcp $\epsilon$ phase transformation around $13$ GPa \cite{Jensen2009}. The geometrical mapping from the bcc to the hcp phase has been an active area of study for decades with several phase transition mechanisms proposed in the literature \cite{burgers1934, Mao1978, rong1984, potter1973}. In our study, XRD data provides information on microstructural changes in Fe across the $\alpha$$\rightarrow$$\epsilon$ phase boundary.~A forward diffraction and texture analysis approach is employed to relate measured azimuthal ($\phi$) diffraction intensities with those predicted from mechanistic transformation pathway models. 

Texture measurement using X-ray diffraction usually involves measuring pole figures and inverting the pole figures to obtain the full orientation distribution function. While pole figure measurements have been conducted for \changen{static compression experiments in iron \cite{Merkel2020,Freville2023},} and used to measure the orientation relationship between ambient and high-pressure phases, due to the \textit{ns} timescales of shock compression experiments, the measurement of pole figures with reasonable angular coverage is not feasible in our experiments.~Only a single ring of the complete pole figure is measured, which makes crystallographic texture determination using pole figure inversion ill-conditioned. \change{This problem can be further exacerbated by peak overlap of the low-pressure and high-pressure peaks}. Therefore, we employ a forward diffraction model to measure the orientation relationship (OR). Given all experiment parameters, the model computes the intensity distribution in $2\theta-\phi$ space. These include:
\begin{enumerate}
	\item The crystal structure, lattice parameters, and phase fraction of the ambient and high-pressure phase, including the Debye-Waller factors. This information is used to compute the powder diffraction intensity in $2\theta$.
	\item Crystallographic texture of the ambient and high-pressure phase. This information modulates the powder intensity in the $\phi$ direction.
	\item X-ray source specification such as polarization and meaningful peak shapes for the X-ray source. In our case, we use the ``pink" beam function, which is a convolution of back-to-back exponential functions with a Gaussian and Lorentzian function described in Ref.~\cite{vondreele2021}.
\end{enumerate}
The crystallographic texture is represented using a finite element representation of the Rodrigues space fundamental zone \cite{Barton2002,Bernier2006,Singh2020}. 

\begin{figure}[!t]
	\begin{center}
		\includegraphics[width=\columnwidth]{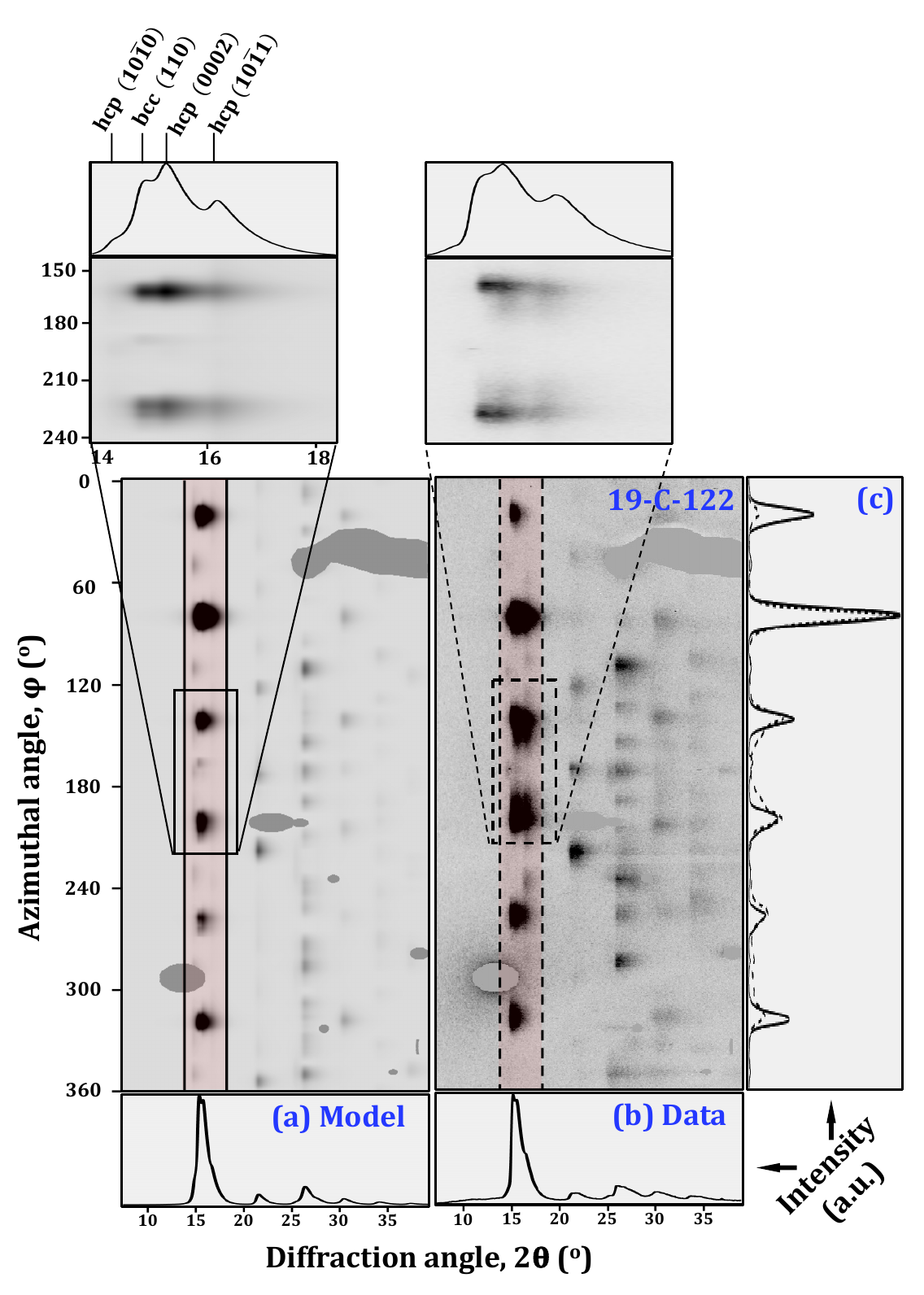}
		\caption{\textbf{$\vert$ Texture as a result of Burger's OR.} \textbf{a.} Forward model simulation of XRD pattern and \textbf{b.} experimentally measured XRD pattern for Fe shock compressed to 25(7) GPa (shot 19-C-122).
			The diffraction data is a combination of the ambient $\alpha$-phase (ahead of the shock wave) and the compressed $\epsilon$-phase (e.g., see top panels). The orientation distribution of the high-pressure phase was determined by applying Burger's OR to the ambient $\alpha-$phase orientation distribution. We report the averaged profiles in 2-$\theta$ in lower panels and profiles averaged in $\phi$ over the regions of interest defined by the solid (model) and dashed boxes (data) in \textbf{c}. \change{The readers are referred to SUpp. Mat. Fig.~\ref{fig:texture_other} for a complete collection so experimental and simulated diffraction from ambient Fe and the other ORs listed in Table~\ref{Table:OR_summ}.}
		}
		\label{fig:Texture}
	\end{center}
	\squeezeup
	\squeezeup
\end{figure}

The samples used in lower pressure shots (P $< 60$ GPa, see Table~\ref{table:Summ_table}) were highly textured and had significant intensity variation in the azimuthal direction of the Debye-Scherrer rings (see Supp. Mat. Fig.~\ref{fig:texture_other}\textbf{a}). Upon compression, the high-pressure hcp $\epsilon-$phase had a repeatable intensity distribution \change{with respect to} the ambient pressure texture. We used the texture to compute the complete pole figures using the methodology described in Ref.~\cite{Barton2002}. We show the pole figures computed for the ambient bcc phase using the forward model in Fig.~\ref{fig:bcc_calc_pfs}. We measure only a single ring of the entire pole figure in the experiment (shown by the red dotted circle in that figure). The intensity variation along this ring modulates the powder diffraction intensity for each diffraction line. We show the resulting diffraction pattern after applying the azimuthal modulation in Fig.~\ref{fig:Texture}c, \change{(see also Fig.~\ref{fig:texture_other}a for other previously reported ORs).} 

A static compression study of single crystal Fe unambiguously demonstrated that the OR between the single crystal $\alpha$ and high-pressure $\epsilon-$phase was the Burger's OR [$(110)_{\alpha} \vert\vert (0001)_{\epsilon}$ and $[1\bar{1}1]_{\alpha} \vert\vert [11\bar{2}0]_{\epsilon}$] \cite{Dewaele2015}. The authors observed all $12$ orientation variants predicted by Burger's mechanism. However, the mechanism during dynamic compression experiments is still unknown. An alternative mechanism has been proposed in Ref.~\cite{Mao1978} resulting in the OR $(110)_{\alpha} \vert\vert (0001)_{\epsilon}$ and $[001]_{\alpha} \vert\vert [2\bar{1}\bar{1}0]_{\epsilon}$. We refer to this as the Mao-Bassett-Takahashi (MBT) OR. The MBT mechanism results in $6$ orientation variants that can form from a single $\alpha-$phase grain, half the number of variants formed in the Burger's OR \cite{burgers1934}. We also considered experimentally observed bcc-hcp ORs for other material systems \cite{potter1973,rong1984}. We show the distribution of some low-index $\epsilon-$phase planes formed as a result of these ORs in Fig.~\ref{fig:pf_ors}a-d. Three of the four ORs produce very similar distributions of the crystallographic planes. The $\{0001\}_{\epsilon}$ planes formed by Burger's and MBT OR are identically distributed, while the angular separation of the $\{11\bar{2}0\}_{\epsilon}$ plane is only $5.26^{\circ}$. The OR observed by Potter \cite{potter1973} produces twice the number of variants as the Burger's OR, with each pair of the variants separated by $\sim$$2^{\circ}$ to the Burger's variant. We list the full ORs in Table~\ref{Table:OR_summ}.
We show the pole figure of some low-index $\epsilon-$phase planes formed due to these ORs in Supp. Mat. Fig.~\ref{fig:pf_ors}. Note that these pole figures are for the cube orientation of the ambient bcc phase such that $(001)_{\alpha} \vert\vert \textbf{z}_{\textrm{lab}}$ and $[100]_{\epsilon} \vert\vert \textbf{x}_{\textrm{lab}}$. Since the ambient $\alpha-$phase has a pole figure different from the cube orientation as shown in Fig.~\ref{fig:bcc_calc_pfs}, the pole figures for the high-pressure $\epsilon-$phase formed as a result of the different ORs listed in Table.~\ref{Table:OR_summ} is calculated by applying the starting orientation of the ambient bcc phase to the pole figures shown in Fig.~\ref{fig:pf_ors}. We present the calculated pole figures for the high-pressure hcp phase following the Burger's, Mao-Bassett-Takahashi, Potter, and Rong-Dunlop orientation relationships in Figs.~\ref{fig:bor_calc_pfs}, \ref{fig:mbt_calc_pfs}, \ref{fig:potter_calc_pfs}, \ref{fig:rong_calc_pfs} respectively, and the resulting diffraction pattern in Figs.~\ref{fig:Texture}d, \ref{fig:texture_other}b-d, respectively.

We report the results of the forward calculation and experimental data for shot 19-C-122 [25(7) GPa] in Fig.~\ref{fig:Texture}a,b respectively. We determined the lattice parameters by performing a LeBail fit to the 1-D lineouts (see Figs.~\ref{fig:Lineout_1}, \ref{fig:Lineout_2}, \ref{fig:Lineout_3}). Here, we applied the Burger's OR to the ambient phase texture to obtain the $\epsilon-$phase texture, with good agreement to the experimentally measured XRD pattern. The diffraction signal consists of ambient $\alpha-$ ($20 \%$) and high-pressure $\epsilon-$phase ($80 \%$), with a complete transformation to all variants of the $\epsilon-$phase.

We compare the results of the forward model with the experimental measurements by comparing the azimuthal variation of some lower-angle diffraction lines. While we rule out the Rong-Dunlop OR \cite{rong1984} based on this calculation, our data are unable to distinguish between the ORs from Burger \cite{burgers1934}, Mao-Bassett-Takahashi \cite{Mao1967}, or the Potter mechanisms \cite{potter1973}. \change{Our analysis shows that, for future experiments, high-quality single crystal samples (having a small orientation spread), coupled with highly monochromatic X-rays are needed to unambiguously identify the phase transition OR for iron during dynamic compression.}

\subsection{\changen{Grain size refinement}}
In a shock compression study of single crystal iron along the $[100]$ direction \cite{hawreliak2008}, the authors reported formation of polycrystalline hcp $\epsilon-$phase with the grain size between $2$ and $15$ nm, with the conclusion that ``single-crystal iron becomes nanocrystalline in shock transforming from $\alpha$ to $\epsilon$ phase." While our \textit{in-situ} diffraction results do not disagree with these results, it presents a more nuanced picture of the grain refinement. The overall texture of the ambient $\alpha$ and the high-pressure $\epsilon-$phase is explained well by a unimodal orientation distribution. This distribution function is akin to a Gaussian distribution around specific orientations. The full width at half-max (fwhm) for the ambient and high-pressure phase that gives the best agreement with experimental data is $\sim 2^{\circ}$ and $\sim 4^{\circ}$, respectively. While each grain retains the mean orientation predicted by the phase transformation mechanism, the orientation distribution within each grain points to microstructure refinement with the possibility that each ``sub-grain" within that grain is nm-sized. \change{There is, however, no large-scale grain refinement/rotation associated with the $\alpha$--$\epsilon$ transformation and/or subsequent plastic deformation within the $\epsilon$-phase that would result, in our experiments, as an almost powder-like diffraction pattern from the high-pressure $\epsilon-$phase.}

\medskip

\subsection{\changen{Hugoniot measurements}}
\begin{figure}[t!]
	\centering
	\includegraphics[width=1\columnwidth]{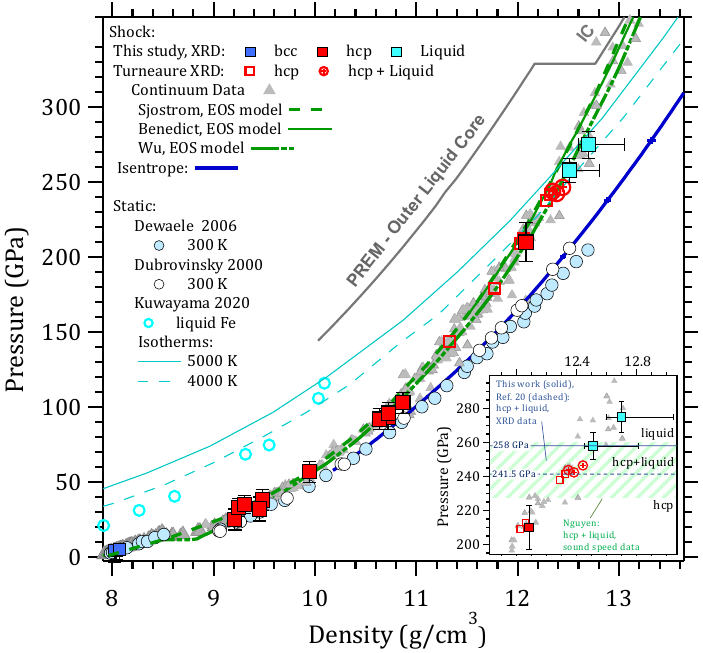}
	\caption{\textbf{Pressure versus density.} 
		$P$-$\rho$ data from this study (bcc - blue squares, hcp - red squares, and liquid - cyan squares), and a previous XRD study \cite{Turneaure2020} (hcp - red open squares, hcp+liquid - red crossed circles). Previously published Hugoniot \cite{Smith2018} (gray triangles), and 300 K static data \cite{Dewaele2006,dubrovinsky2000b,Kuwayama2020} (blue and white filled circles) are also shown, as well as an isentrope determined from laser ramp-compression experiments (blue curve) \cite{Smith2018}. High-temperature liquid static data is shown by the open cyan circle along with calculated 4000 K and 5000 K isotherms \cite{Kuwayama2020}. Three Hugoniot models are also shown (green curves) \cite{Wu2023, benedict2022, sjostrom2018}. (Inset) A magnified view shows agreement \textcolor{black}{for melt completion pressures along the Hugoniot between our XRD data and previous sound speed measurements \cite{nguyen2004}. However, previous XRD data \cite{Turneaure2020} and sound speed measurements \cite{nguyen2004} disagree on the melt onset pressures along the Hugoniot. These combined XRD studies constrain the hcp+liquid coexistence along the Hugoniot between 241.5(3)-258(8) GPa, whereas it is 225(3)-258(8) GPa using the combined sound speed measurements and XRD data from this study.}}
	\label{fig:Prho}
	\squeezeup
	\squeezeup
\end{figure}

\begin{figure}[!ht]
	\begin{center}
		\includegraphics[width=0.96\columnwidth]{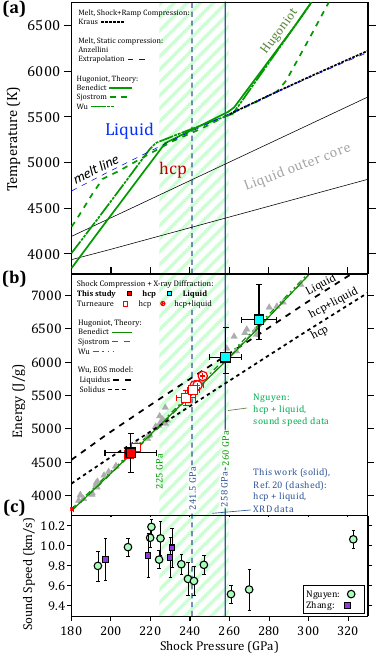}
		\caption{\textbf{Fe pressure-temperature phase map. a.} High $P$-$T$ region of the Fe phase map, which shows the stability regions for the $\epsilon$-hcp, and the liquid phases.~The melt line is constrained by static \cite{anzellini2013}, and nanosecond timescale shock+ramp compression experimental data \cite{Kraus2022}. Three theoretical bounds on the Hugoniot are represented by the solid \cite{benedict2022} and dashed \cite{sjostrom2018}, and dashed-dotted green curves \cite{Wu2023}. The extent of solid-liquid coexistence in the Benedict \cite{benedict2022} and Wu \cite{Wu2023} models are set by the measured changes in sound speed under shock compression of Nguyen and Holmes \cite{nguyen2004} (see \textbf{c.} and Fig.~\ref{fig:Nguyen}), which are inferred in that study to reflect the onset and completion of melt. \textcolor{black}{Our XRD data point for melt completion at 258(8) GPa is consistent with previous sound speed determinations of solid-liquid coexistence (green band). However, the XRD data point for melt onset reported in \cite{Turneaure2020} at 241.5(3) GPa is inconsistent with previous measurements. This may be related to the insensitivity of X-ray diffraction to measure small volumes of liquid (see Supplementary Materials).} Also shown is the range of possible temperatures for the earth's outer core \cite{Dziewonski1981}. A more inclusive representation of experimental and theoretical studies of Fe in $P$-$T$ space is shown in Fig. \ref{fig:PT_full}. \textbf{b.} Energy vs. pressure plot, which shows our shock-compression data, previous Hugoniot data (\cite{Smith2018}, grey triangles), theoretical predictions for the Hugoniot \cite{Wu2023, sjostrom2018, benedict2022}, and solidus and liquidus curves (derived from EOS in Ref. \cite{Wu2023}). \textbf{c.} Sound speed data from \changen{Refs.~\cite{nguyen2004,zhang2023}}}.
		\label{fig:PT_3}
	\end{center}
	\squeezeup
	\squeezeup
	\squeezeup
\end{figure}

Figure~\ref{fig:Prho} shows our data in pressure-density space, color-coded with measured phase (blue: bcc, red: hcp, cyan: liquid), and plotted against other shock, ramp, and static compression datasets.~Also plotted is the $P$-$\rho$ curve estimated from the Preliminary Reference Earth Model (PREM) \cite{Dziewonski1981} for the outer liquid and solid inner core (IC). At 180 GPa, where the $P$-$T$ conditions along the Fe Hugoniot \cite{benedict2022} are comparable to those found within the outer liquid core ($\sim$180 GPa, $\sim$4080 K, see Fig. \ref{fig:PT_3}), there is a 10.1\% difference in the density. The lower density of the Fe-rich outer core is expected due to the inclusion of 2.7-11.5\% by weight of low-Z impurities \cite{Hirose2021}. The inset to Fig.~\ref{fig:Prho} shows the magnified view of the structural measurements from this study (solid horizontal line). The green hatched region shows the coexistence region determined using sound speed measurements from a previous study \cite{nguyen2004}.~A dashed horizontal line also shows the melt onset pressure from a recent XRD study in Ref.~\cite{Turneaure2020} at $241.5(3)$ GPa. {\color{black} Based on the XRD study presented in this work (melt completion) and in the XRD study of Ref. \cite{Turneaure2020} (melt onset), the liquid-solid coexistence lies between 241.5(3)-258(8) GPa.} This coexistence is smaller than the 225(3)-260(3) GPa range inferred by Nguyen and Holmes \cite{nguyen2004} from experiments that measure changes in sound speed as a function of pressure for microsecond shock compression of mm-thick Fe samples. \changen{A more recent sound speed measurement by Zhang et al. \cite{zhang2023} has shown no discernible drop in sound speed around the $225$ GPa pressure inferred as the melt onset from the Nguyen and Holmes measurement (see Supplementary Materials Fig.~\ref{fig:Nguyen}). We note that a reanalysis of the Nguyen and Holmes \cite{nguyen2004} data is planned based on a reevaluation of uncertainty analysis \cite{akin2015}.}

One contributing factor for the discrepancy in melt onset pressures between the two experimental approaches (XRD and sound speed) \emph{may} be due to a lower sensitivity of XRD in detecting the weak diffuse signal of the incipient liquid phase. 
\change{With the sound speed technique 
	both the shear and bulk moduli contribute to the sound velocity in a solid. Upon melting, the shear modulus $\rightarrow$ 0, resulting in a sharp decrease in sound velocity for pressures where the Hugoniot intersects with the melt curve \cite{nguyen2004} (see Fig. \ref{fig:Nguyen}).~We expect this measured change in sound speed} to be more sensitive than XRD measurements to the onset of small volumes of liquid (see Supplementary Methods).
~However, because of the sparsity of data points in the measurements presented in Ref.~\cite{nguyen2004} (no data between 247 and 260 GPa, see Fig. \ref{fig:PT_3}c), the sound speed measurements are less constraining for the completion of the melt.

The completion of melt, however, is well constrained \change{within} XRD  as small volumes of a textured solid are easily observed. We note that while the lowest pressure liquid-only data presented here is at 258(8) GPa, the pressure for melt completion is constrained by this value and the observation of high-pressure hcp from a previous XRD study \cite{Turneaure2020}, to be in the range 246.5(1.8)-258(8) GPa.

The consistency in the pressures for melt completion between experiments with shock compression durations, \change{in the $\mu s$ timescale for gas-gun experiments and in the $ns$ timescale for laser-shock experiments,} differ by over three orders of magnitude. This observation is consistent with a lack of time-dependence in the melt of Fe.~This is in contrast to the reported strong time-dependence reported for the low-pressure $\alpha$$\rightarrow$$\epsilon$ phase transformation \cite{Smith2013}. 

Figure~\ref{fig:PT_3}a shows the pressure-temperature ($P$-$T$) phase map of Fe in the vicinity of the melt line with three recent Hugoniot models \cite{sjostrom2018, benedict2022, Wu2023}. These hugoniot models were constructed to cross the melt line in agreement with the extrapolated melt of the static compression study of Anzellini \emph{et al.} \cite{anzellini2013} and the dynamic ns-compression work of Kraus \emph{et al.} \cite{Kraus2022}. The pressure extent of the solid-liquid coexistence for the Wu \emph{et al.} \cite{Wu2023}, and Benedict \emph{et al.}~\cite{benedict2022} Hugoniots were set to agree with the 225(3)-260(3) GPa range inferred from sound speed measurements from Nguyen and Holmes \cite{nguyen2004} (see also Supplementary Materials Fig. \ref{fig:Nguyen}). 

We plot the internal energy of the solidus and liquidus curves from the Wu EOS model in Fig. \ref{fig:PT_3}b \cite{Wu2023}. By using the Rankine-Hugoniot equations, based on conservation of energy, \change{and assuming zero reference pressure and energy state, i.e., $P_0, E_0 = 0$, $E = \frac{1}{2} P\left(\rho-\rho_0\right)/(\rho\rho_0)$} \cite{zel2002}, where P is the experimentally-determined shock pressure, $\rho$ is the density from X-ray diffraction measurements, and $\rho_0$ is the initial density (7.874 g/cm$^3$), we can compare our measured Hugoniot data with the calculated Hugoniot and melting curve, and provide a valuable description of the Fe phase diagram in pressure-energy space (Fig.~\ref{fig:PT_3}b). 

\medskip
\section{Discussion}
\squeezeup
Latent heat release ($\Delta$H$_{m}$) associated with freezing at the earth's inner core boundary (ICB) is regarded as one of the primary drivers ($\sim$20\% of total energy contribution \cite{LABROSSE2015}) of convection currents in the outer core. This value has not been directly constrained experimentally at ICB conditions, and therefore, models describing the core dynamics rely on theoretical calculations of $\Delta$H$_{m}$ \cite{sun2018,cuong2021, sun2022}. \textcolor{black}{The entropy of fusion ($\Delta S_{m} = \Delta$H$_{m}/T_{m}$) for simple systems is expected to approach the gas constant, $\textrm{R}$ asymptotically at high pressure \cite{Stishov1975}, commonly referred to as Richard's rule. This behavior is expected since our liquid structure measurement of Fe on the Hugoniot demonstrates that Fe remains a simple liquid at these conditions.}

~Experimentally, this value can be constrained at the high $P$-$T$ conditions accessed along the Hugoniot by a measure of the extent of solid-liquid coexistence and knowledge of the equation of state of Fe (see Supplementary Materials). The measurement of melt completion in this study, together with previous measurements of melt onset \cite{nguyen2004}, can be used to constrain the latent heat of fusion at 258 GPa to 623 J/g, equivalent to the entropy of fusion, \change{$\Delta S_{m} = 0.8\textrm{R}$}.~This value is lower than the expected value of $\textrm{R}$, and \change{is approximately half the value reported} in three recent theory predictions at this pressure \cite{cuong2021}, and suggests that the energy contribution to the geodynamo budget due to the freezing of Fe is smaller than previously thought. \change{We also note that the latent heat at the same pressure derived from the recently published Fe EOS model of Wu \emph{et al.} \cite{Wu2023} is 666.9 J/g, corresponding to $\Delta S_{m} = 0.82\textrm{R}$. This value is very close to our reported value since this EOS model used the sound speed data of Nguyen et al. \cite{nguyen2004} as a constraint, which is consistent with our study.}


Assuming that the latent heat of fusion at the inner core boundary pressure (330 GPa) changes negligibly due to higher pressure and a small amount of impurities, we can compute the total power output due to the solidification of the inner core. This quantity is the ratio of the heat generated due to the solidification and the time over which the solidification occurred. The heat generated due to solidification is given by the product of $\Delta \textrm{H}_{m}$ of pure iron and the mass of the inner core ($1.1 \times 10^{23}$ Kg \cite{SOROKHTIN2011}). The latest estimates for the age of earth's inner core range between $0.565 - 4$ billion years \cite{LABROSSE1997,stacey2008,Bono2019,Zhang2020}. When considering our estimates of latent heat estimates, this results in power output of \change{$0.54 - 3.85$} TW, which is lower than previous estimates \cite{Anderson1997}. This discrepancy suggests the importance of compositional convection and/or radioactive heating \cite{Kanani2003,Gando2011} in maintaining the earth's geodynamo.

The depression in the melting point over pure material can be estimated, assuming an ideal mixing model and thermodynamic equilibrium between pure solid and liquid with impurities. It is given by \cite{Anderson1997}
\change{
	\begin{equation}
		{T} = \frac{T^{*}}{1 - \left(R T^{*}/\Delta\textrm{H}_m\right)\log(1-\chi_{s})}.
		\label{eq:Tm}
	\end{equation}
}
Here, $\chi_{s}$ is the mole fraction of impurities, $R$ is the gas constant, $T^{*}$ is the melt temperature of the pure material, and $T$ is the melt temperature in the presence of impurities. For the same impurity level, a smaller latent heat will result in a larger drop in the melting temperature. The liquid outer core is estimated to contain approximately $2.7 - 11.5\%$ light impurities by weight \cite{Hirose2021}. The melting temperature at the inner core boundary (ICB; P = 330 GPa)  using the melt curve in Ref.~\cite{Kraus2022} is approximately $6220$ K. \textcolor{black}{Using the estimate of the latent heat in this study, the temperature of outer core side of ICB is constrained to a range of \change{$3850 - 5650$} K. We note that the upper bound of this range is more accurate than the lower bound as Eqn.~\ref{eq:Tm} holds better for dilute solutions, i.e., small $\chi_{s}$.}

\section{Conclusions}
\squeezeup
In conclusion, we present a detailed structural study of Fe under shock compression from 25 to 275 GPa. We analyzed liquid scattering data to obtain the scattering factors, the radial distribution function, and the density on the shock Hugoniot. We found the upper bound shock pressure for complete melt to be 258(8) GPa. We utilized a forward diffraction model for texture analysis of the ambient bcc $\alpha-$phase and high-pressure hcp $\epsilon-$phase. We tested some commonly observed orientation relationships between the bcc and hcp phases for consistency with our diffraction data. We can rule out the Rong-Dunlop OR. However, our data can not distinguish between Burger's, MBT, and Potter's orientation relationship. Our XRD data, along with previous melt onset determination using sound speed measurements \cite{nguyen2004}, suggest an hcp-liquid coexistence along the Hugoniot  (225(3)-258(8) GPa). Our data is consistent with a \change{maximum} internal energy change due to melting of \change{623} J/g at 258(8) GPa, which is approximately half of recent theoretical studies and a value that suggests that the energy contribution to the geodynamo budget due to freezing of Fe at the inner core boundary is smaller than previously thought. The small latent heat also suggests a larger decrease in the melting temperature in the presence of impurities than previously estimated. \changen{We note recent measurements by Zhang et al. \cite{zhang2023} indicate no first-order phase transition in iron on the Hugoniot up to 230.8(1) GPa. The new data suggests that our calculated latent heat is likely an upper bound.}

\section*{Acknowledgements}
\squeezeup
We want to thank Carol Ann Davis for her help in preparing the Fe targets.~Paulo Rigg, Pinaki Das, Ray Gunawidjaja, Yuelin Li, Adam Schuman, Nicholas Sinclair, Xiaoming Wang, and Jun Zhang at the Dynamic Compression Sector are gratefully acknowledged for their expert assistance with the laser experiments.~We thank Yoshi Toyoda for his assistance with VISAR measurements during the laser experiments, and Richard Kraus for helpful discussions on pressure uncertainies.~This work was performed under the auspices of the US Department of Energy by Lawrence Livermore National Laboratory under contract number DE-AC52-07NA27344.~This publication is based upon work performed at the Dynamic Compression Sector, which is operated by Washington State University under the US Department of Energy (DOE)/National Nuclear Security Administration under Award No. DE-NA0003957. This research used resources from the Advanced Photon Source, a DOE Office of Science User Facility operated for the DOE Office of Science by Argonne National Laboratory under Contract No. DE-AC02-06CH11357.

\bibliographystyle{elsarticle-num}
\bibliography{references}
\onecolumn
\noindent
\begin{center}
\Large{\textbf{A structural study of hcp and liquid iron 
\\ under shock compression up to 275 GPa}}
\end{center}
\bigskip
\noindent
\normalsize{Saransh Singh, Richard Briggs, Martin G. Gorman, Lorin X. Benedict, Christine J. Wu, Sebastien Hamel, Amy L. Coleman, Federica Coppari, Amalia Fernandez-Pa$\tilde{n}$ella, Christopher McGuire, Melissa Sims, June K. Wicks, Jon H. Eggert, Dayne E. Fratanduono,  Raymond F. Smith}

\bigskip
\bigskip
\noindent
\Large{Supplementary Materials}

\renewcommand{\thesection}{S\arabic{section}}
\renewcommand{\thefigure}{S\arabic{figure}}
\renewcommand{\thetable}{S\arabic{table}}

\setcounter{section}{0}
\setcounter{figure}{0}
\setcounter{table}{0}

\normalsize 

\bigskip
\noindent
\underline{\textbf{Supplementary Material Tables:}}

\bigskip
\noindent
\textbf{Table \ref{table:Summ_table}:} Summary of data.



\bigskip
\noindent
\underline{\textbf{Supplementary Material Figures:}}

\begin{figure}[h!]
\begin{center}
\includegraphics[width=0.8\columnwidth]{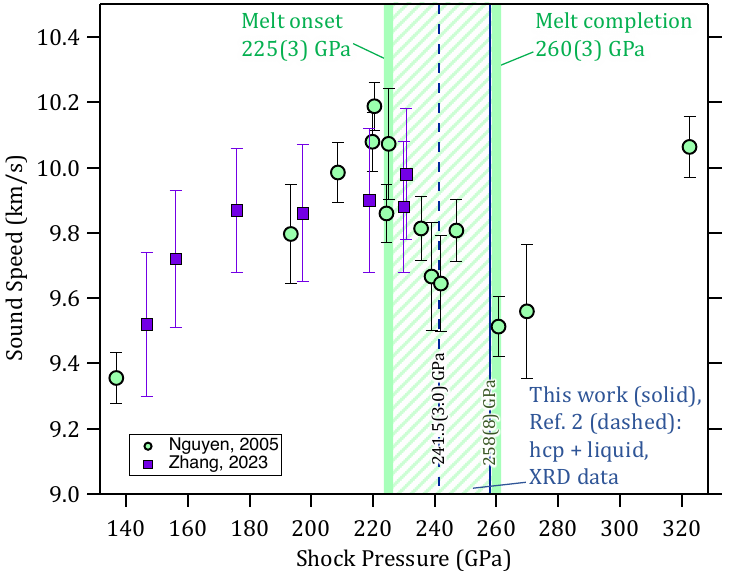}
\caption{\textbf{Pressure versus sound speed.} Sound speed data from Nguyen and Holmes \protect\cite{nguyen2004cp} (green symbols) and Zhang \emph{et al.} \cite{zhang2023cp} (purple squares) as a function of shock pressure. We note that the drop in sound speed observed by Nguyen and Holmes was interpreted as the onset of melt at 225$\pm$3 GPa. However the recently published data from Zhang up to 230.8$\pm$1 GPa, using the same experimental technique as Nguyen and Holmes, did not show the same sound speed drop. While these data provide no direct structural information, changes in sound speed are interpreted as the onset and completion of melt, as indicated. In those experiments, the shock duration applied to mm-thick Fe samples was $\sim\mu$s. In contrast the combined XRD and laser shock compression experiments reported here use nanosecond shock compression of 21 $\mu$m thick samples. Our lowest pressure liquid only point is at 258(8) GPa (solid vertical line). This data combined with observations from a previous XRD study \cite{Turneaure2020cp} places the completion of shock melt in the range 246.5(1.8)-258(8) GPa. The hcp+liquid coexistence region from our data and an earlier XRD study \protect\cite{Turneaure2020cp} are consistent with with onset (vertical dashed line) and completion of melt of 241.5(3.0) GPa and 258(8) GPa, respectively.  
}
\label{fig:Nguyen}
\end{center}
\squeezeup
\squeezeup
\end{figure}

\bigskip
\noindent
\textbf{Figure \ref{fig:Nguyen}:} hcp+liquid coexistence.

\begin{figure}[!h]
\begin{center}
\includegraphics[width=0.8\columnwidth]{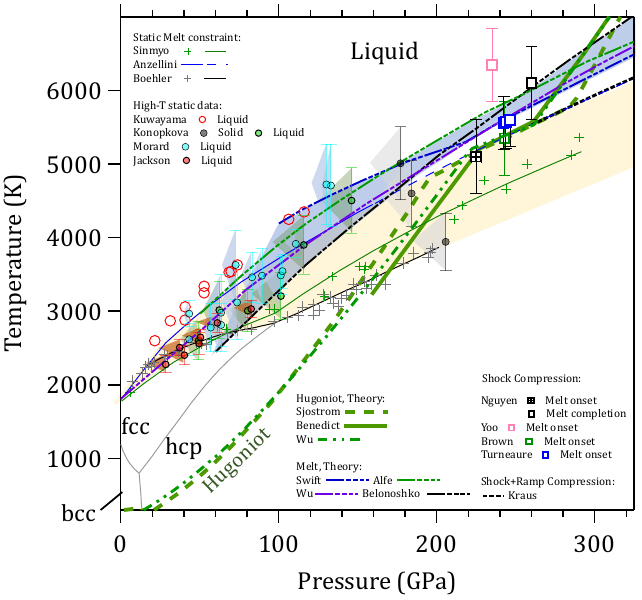}
\caption{\textbf{Temperature versus Pressure.} Pressure-temperature phase map of Fe with three solid phases and a liquid phase. The melt line as constrained by static compression studies is bound by yellow shaded region \protect\cite{anzellini2013cp,boehler1990cp,boehler1993cp,Sinmyo2019cp,konopkova2021, morard2018,jackson2013}. Bounds provided by theoretical calculations are shown by the blue shaded region \protect\cite{alfe2009cp,swift2020cp,belonoshko2000cp,wu2011}. Three theoretical bounds on the Hugoniot are represented by the green curves \protect\cite{Wu2023cp,sjostrom2018cp,benedict2022cp}. Determined pressure and estimated temperatures for previous sound speed measurements under shock compression (no direct lattice level observations) for the onset of shock melt (colored squares \protect\cite{nguyen2004cp,Brown1986cp,yoo1993}) and full melt (white square \protect\cite{nguyen2004cp}) are also shown, as well as the calculated melt onset temperature from a recent laser shock + XRD study \protect\cite{Turneaure2020cp}. 
}
\label{fig:PT_full}
\end{center}
\squeezeup
\squeezeup
\end{figure}

\noindent
\textbf{Figure \ref{fig:PT_full}:} Temperature versus Pressure map for Fe.

\noindent
\change{\textbf{Figure \ref{fig:PT_melt}:} Melt curve and Hugoniot curve from updated equation of state.}



\begin{figure}[h!]
	\begin{center}
	\includegraphics[width=0.5\columnwidth]{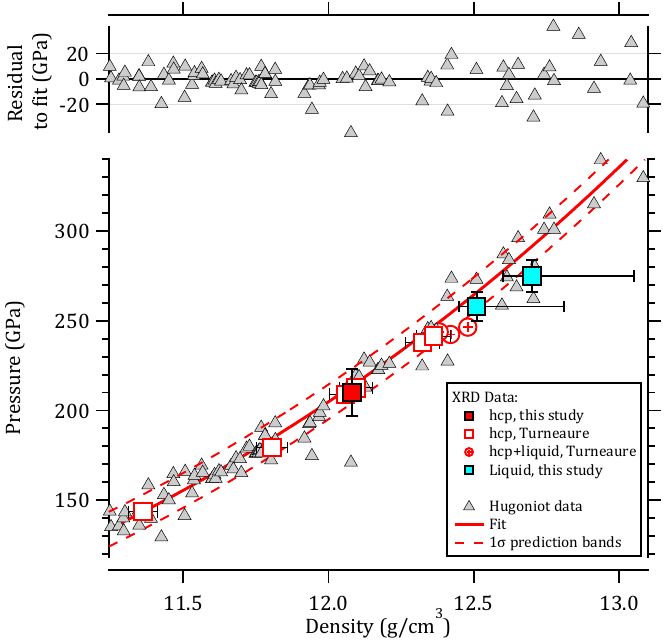}
	\caption{\textbf{Hugoniot Fit.} The red curve is based on a fit to previous Hugoniot data (grey triangles), and the dashed black lines are $1\sigma$ prediction bands which represents the bounds within which there is an expectation 68.3\% of shock data points should fall within. The upper plots shows residuals to the fit. 
	}
	\label{fig:Hugoniot_Fit}
	\end{center}
	\squeezeup
	\squeezeup
\end{figure}

\noindent
\textbf{Figure \ref{fig:Hugoniot_Fit}:} Fit to Hugoniot data.

\begin{figure}[h!]
	\begin{center}
	\includegraphics[width=1\columnwidth]{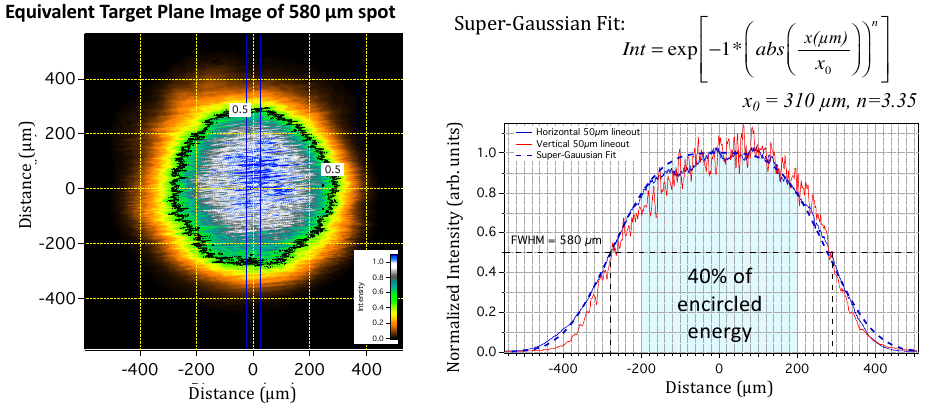}
	\caption{\textbf{DCS laser spot.} (left) Equivalent target plane image of DCS 580 µm diameter laser spot with a normalized intensity scale. The 0.5 intensity contour is highlighted. (right) Normalized intensity as a function of distance for 50 $\mu$m wide vertical (red) and horizontal (blue) lineouts from the image on the left (vertical lineout averaged between blue line in left image). The intensity lineouts are described by a super-Gaussian spatial profile (dashed blue curve), as indicated. Within this distribution 40\% of the total energy is contained within the central 400-$\mu$m diameter spot (blue shaded region).
	}
	\label{fig:DCS}
	\end{center}
	\squeezeup
	\squeezeup
\end{figure}

\medskip
\noindent
\textbf{Figure \ref{fig:DCS}:} Laser focal spot spatial profile.

\medskip
\noindent
\textbf{Figure \ref{fig:VISAR}:} VISAR traces and Laser power profiles.

\noindent
\textbf{Figure \ref{fig:pf_ors}:} Pole figures for some low index planes of the hexagonal closed packed phase as a result of the different phase transition mechanisms.

\noindent
\textbf{Figure \ref{fig:bcc_calc_pfs}:} Calculated pole figures for the ambient bcc-phase

\noindent
\textbf{Figure \ref{fig:bor_calc_pfs}:} Calculated pole figures for the high pressure hcp-phase following the Burger's orientation relationship.

\noindent
\textbf{Figure \ref{fig:mbt_calc_pfs}:} Calculated pole figures for the high pressure hcp-phase following the Mao-Bassett-Takahashi orientation relationship.

\noindent
\textbf{Figure \ref{fig:potter_calc_pfs}:} Calculated pole figures for the high pressure hcp-phase following the Potter orientation relationship.

\noindent
\textbf{Figure \ref{fig:rong_calc_pfs}:} Calculated pole figures for the high pressure hcp-phase following the Rong-Dunlop orientation relationship.

\noindent
\textbf{Figure \ref{fig:texture_other}:} Simulated diffraction pattern for (a) ambient $\alpha-$Fe and high-pressure $\epsilon-$phase following the (b) Mao-Bassett-Takahashi, (c) Potter and (d) Rong-Dunlop orientation relationships. The equivalent results for the Burger's orientation relationship reported in Fig.~\ref{fig:Texture}.


\medskip
\noindent
\textbf{Figure \ref{fig:Lineout_1}:} X-ray Diffraction Lineouts (s19-C-122, s19-C-109, s19-C-129, s19-C-128).

\noindent
\textbf{Figure \ref{fig:Lineout_2}:} X-ray Diffraction Lineouts (s19-C-121, s19-C-123, s19-C-124, s19-C-126).

\noindent
\textbf{Figure \ref{fig:Lineout_3}:} X-ray Diffraction Lineouts (s19-C-125, s19-C-127, s19-C-088).

\begin{figure}[]
\begin{center}
\includegraphics[width=0.6\textwidth]{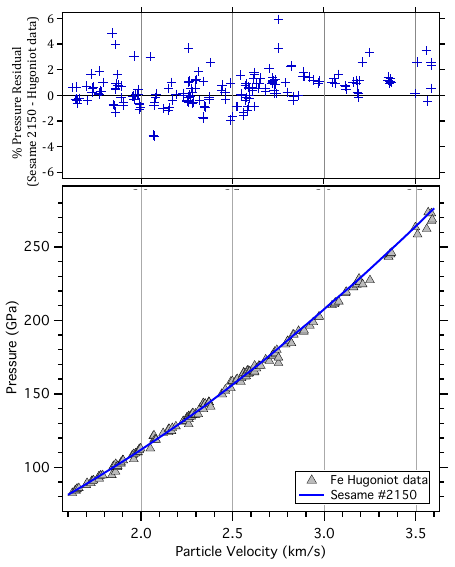}
\caption{\label{fig:2150} Sesame EOS \#2150 for Fe as used in simulation reported in Fig. \ref{fig:Hyades}, compared to previously published Hugoniot data.}
\end{center}
\end{figure}

\noindent
\textbf{Figure \ref{fig:2150}:} Sesame EOS \#2150 for Fe.

\newpage
\change{
\begin{longtable}{|p{0.1\textwidth}|p{0.1\textwidth}| p{0.1\textwidth}|p{0.1\textwidth}|p{0.2\textwidth}|p{0.16\textwidth}|}
\caption{ Summary table of all shots taken in the experiment. } \label{table:Summ_table} \\
\hline
\centering
\textbf{\large Shot} \large $\#$ & \textbf{\large X-ray} \large $\#$ & \textbf{\large Pressure (GPa)} & \textbf{\large Phase} & \textbf{\large Lattice parameters ($\AA$)} & \textbf{\large density (g/cm$^{3}$)}\\
\hline \hline
19-C-122 & 438  & 25(7) & hcp & a$_{\textrm{\tiny{hcp}}}$=2.43, c$_{\textrm{\tiny{hcp}}}$=3.94 & 9.208(5) \\
 \hline
19-C-121 & 432 & 32(8) & bcc + hcp & a$_\text{\tiny bcc}$=2.849 \newline a$_\text{\tiny hcp}$=2.412, c$_\text{\tiny hcp}$=3.984 & 8.023(3) \newline 9.455(4) \\
 \hline
19-C-129 & 496 & 33(6) & bcc + hcp & a$_\text{\tiny bcc}$=2.847 \newline a$_\text{\tiny hcp}$=1, c$_\text{\tiny hcp}$=1 & 8.038(3) \newline 9.242(3)  \\
 \hline
19-C-128 & 489 & 35(6) & hcp & a$_\text{\tiny hcp}$=2.422, c$_\text{\tiny hcp}$=3.926 & 9.301(4) \\
 \hline
19-C-109 & 333 & 38(7) & bcc + hcp & a$_{\text{\tiny bcc}}$=2.847 \newline a$_{\text{\tiny hcp}}$=2.406, c$_{\text{\tiny hcp}}$=3.9 & 8.036(3) \newline 9.482(7) \\
 \hline
19-C-123 & 448 & 57(7) & hcp & a$_\text{\tiny hcp}$=2.376, c$_\text{\tiny hcp}$=3.816 & 9.942(7) \\
 \hline
19-C-124 & 454 & 92(7) & hcp & a$_\text{\tiny hcp}$=2.324, c$_\text{\tiny hcp}$=3.728 & 10.633(5) \\
 \hline
19-C-126 & 468 & 93(8) & hcp & a$_\text{\tiny hcp}$=2.322, c$_\text{\tiny hcp}$=3.702 & 10.72(2) \\
 \hline
19-C-125 & 462 & 96(7) & hcp & a$_\text{\tiny hcp}$=2.312, c$_\text{\tiny hcp}$=3.736 & 10.721(3) \\
 \hline
19-C-127 & 474 & 103(7) & hcp & a$_\text{\tiny hcp}$=2.303, c$_\text{\tiny hcp}$=3.717 & 10.867(4) \\
 \hline
19-C-088 & 182 & 210(13) & hcp & a$_\text{\tiny hcp}$=2.218, c$_\text{\tiny hcp}$=3.602 & 12.081(6) \\
 \hline
19-C-068 & 037 & 258(8) & liquid & -- & 12.51 $\pm$ (0.3, 0.06) \\
 \hline
19-C-117 & 393 & 275(9) & liquid & -- & 12.7 $\pm$~(0.35, 0.1) \\
 \hline
\end{longtable}
}
\clearpage

\clearpage

\section{\change{Updated EOS model of Wu et al.~\cite{Wu2023cp}}}
\begin{figure}[!h]
\begin{center}
\includegraphics[width=0.5\textwidth]{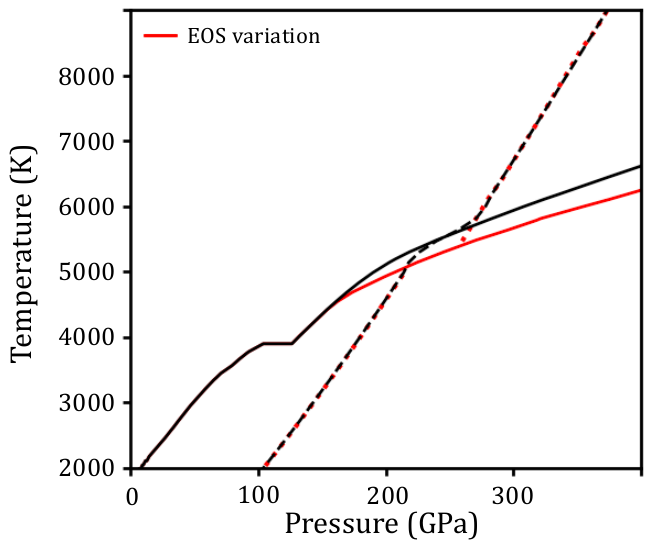}
\caption{\change{\textbf{Melt line and Hugoniot curve of the new EOS.} The solid black curve shows the melt curve reported in Ref.~\cite{Wu2023cp}, while the Hugoniot is shown by the dashed black line. The red curves show the shift in the melt boundary and the Hugoniot based on the slightly lower melt pressures measured in this study. T$_{\text{melt}}$ at the melt-completion pressure is reduced by $\sim$275 K. The updated melt curve leads to slightly better agreement with the Sinmyo \emph{et al.}~\cite{Sinmyo2019cp} melt measurements, and worse agreement with the higher-T$_{\text{melt}}$ data of Anzellini \emph{et al.}~\cite{anzellini2013cp}.} 
}
\label{fig:PT_melt}
\end{center}
\squeezeup
\squeezeup
\end{figure} 
\change{The present experimental work has determined the intersection between the Fe principal Hugoniot and the liquidus line to have a pressure of 258(8) GPa. This is slightly lower than the value reported in Nguyen and Holmes \cite{nguyen2004cp} (260(3) GPa). Since the latter value was used as a constraint when constructing the multiphase Fe EOS published recently by Wu et al.~\cite{Wu2023cp}, it is worth asking: Is it possible to modify that EOS model to agree with these newer data, and if so, what might this change to the Fe EOS imply for the phase diagram of Fe? 

We have attempted to answer these questions by making small modifications to the EOS model of Ref.~\cite{Wu2023cp}. The smallness of these changes is mandated by the fact that this EOS model is already in excellent agreement with a host of ambient, static high-pressure, and dynamic high-pressure experimental data, and large modifications would undermine said agreement. 

Figure \ref{fig:PT_melt} shows the melt curve as a function of pressure for the original baseline EOS model of Ref.~\cite{Wu2023cp} (black solid line), and the principal Hugoniot of this model (black dashed line). The red lines (solid, dashed) show the corresponding curves for a variation of the Fe EOS model made specifically for this work, wherein the present value of 258 GPa was assumed for the pressure of intersection between melt curve and liquidus line. To move this intersection point down in pressure by the desired amount, the density-dependent Debye temperatures of hcp and liquid phases were slightly altered, along with the cold curve of the liquid phase (see Ref.~\cite{Wu2023cp} for a discussion of the phase-dependent free energy model forms and their associated parameters).~As apparent from Fig.~\ref{fig:PT_melt}, this results in a lowering of T$_{\text{melt}}$(P), which also causes the intersection between the Hugoniot and the solidus line to move down in pressure by $\sim$2 GPa. 

This newer version of the Fe EOS model fits the P--~$\rho$ principal Hugoniot data in both hcp and liquid phases just as well as the model versions presented in Ref.~\cite{Wu2023cp}. However, the lowering of T$_{\text{melt}}$(P) by roughly 275 K in this pressure range leads to slightly better agreement with the Sinmyo \emph{et al.} melt measurements \cite{Sinmyo2019cp}, and worse agreement with the higher-T$_{\text{melt}}$ data of Anzellini \emph{et al.} \cite{anzellini2013cp} to which the original baseline Fe EOS model was fit~\cite{Wu2023cp}. From these comparisons, we can conclude that such a change to the Fe EOS and phase diagram are not unreasonable, though we are not able to state which scenario (black vs. red in Fig.~\ref{fig:PT_melt}) is more likely to be correct strictly from the perspective of EOS modeling. It is also important to stress that other EOS model modifications which appease agreement with the present work may be possible which lead to smaller (or at least different) changes to T$_{\text{melt}}$(P); further investigation should be done to explore such possibilities.} 

\section{Sensitivity of XRD techniques to measuring small volumes of liquid/solid\label{sec:sensitivity}}
\begin{figure}[]
\begin{center}
\includegraphics[width=0.6\textwidth]{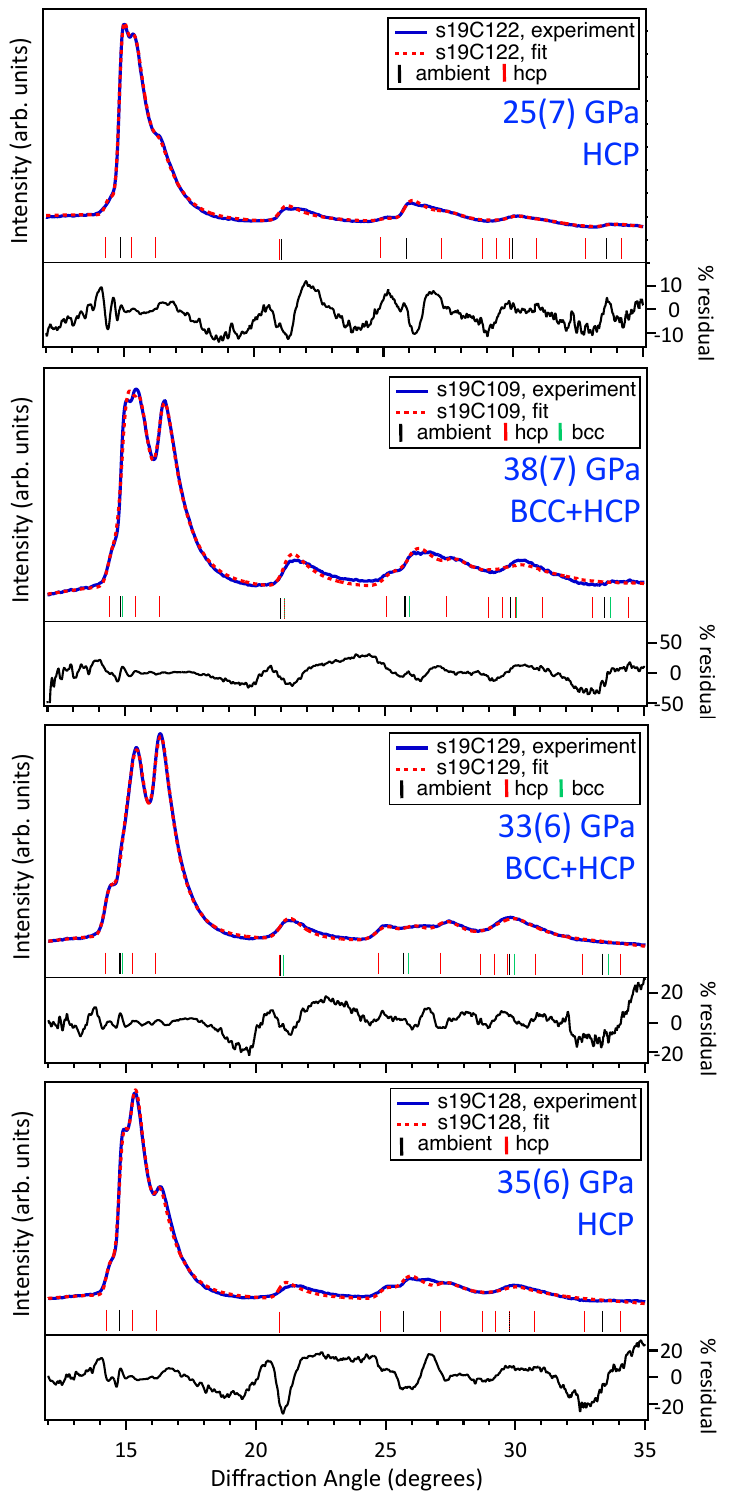}
\caption{\label{fig:Lineout_1} Azimuthally-integrated  profiles for experiments s19C122, s19C109, s19C129, and s19c128 (blue) and fit using the HEXRD software (red dotted) \protect\cite{HEXRDgithubcp}.~All profiles include a contribution from the unshocked-volume (ambient fcc-phase) ahead of the shock front (black ticks), and the shock-compressed fcc-phase (red ticks) and/or bcc-phase (green ticks).~The \% difference between the experimental profile and the fit is shown on the right hand axis.}
\end{center}
\end{figure}

\begin{figure}[]
\begin{center}
\includegraphics[width=0.6\textwidth]{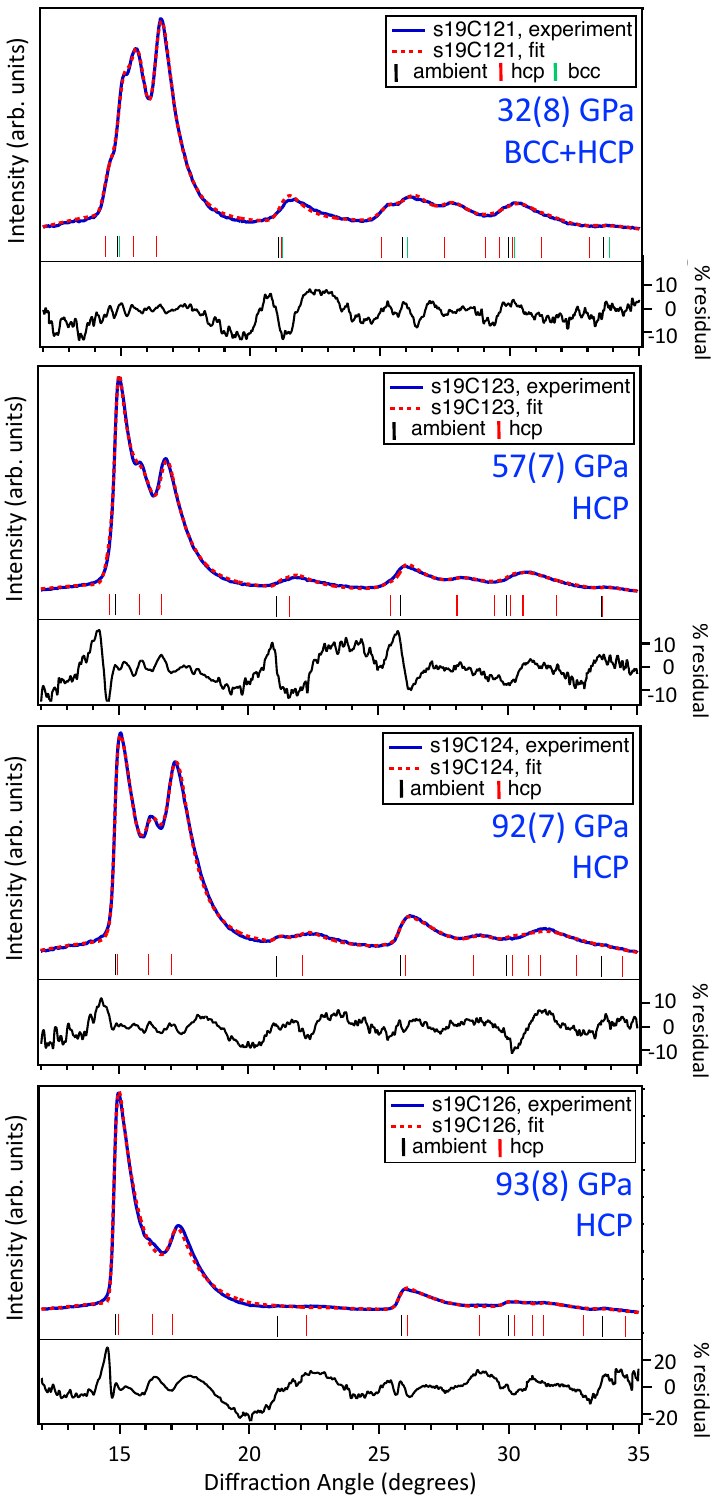}
\caption{\label{fig:Lineout_2} Azimuthally-integrated  profiles for experiments s19C121, s19C123, s19C124, and s19C126 (blue) and fit using the HEXRD software (red dotted) \protect\cite{HEXRDgithubcp}.~All profiles include a contribution from the unshocked-volume (ambient fcc-phase) ahead of the shock front (black ticks), and the shock-compressed fcc-phase (red ticks) and/or bcc-phase (green ticks).~The \% difference between the experimental profile and the fit is shown on the right hand axis.}
\end{center}
\end{figure}

\begin{figure}[]
\begin{center}
\includegraphics[width=0.6\textwidth]{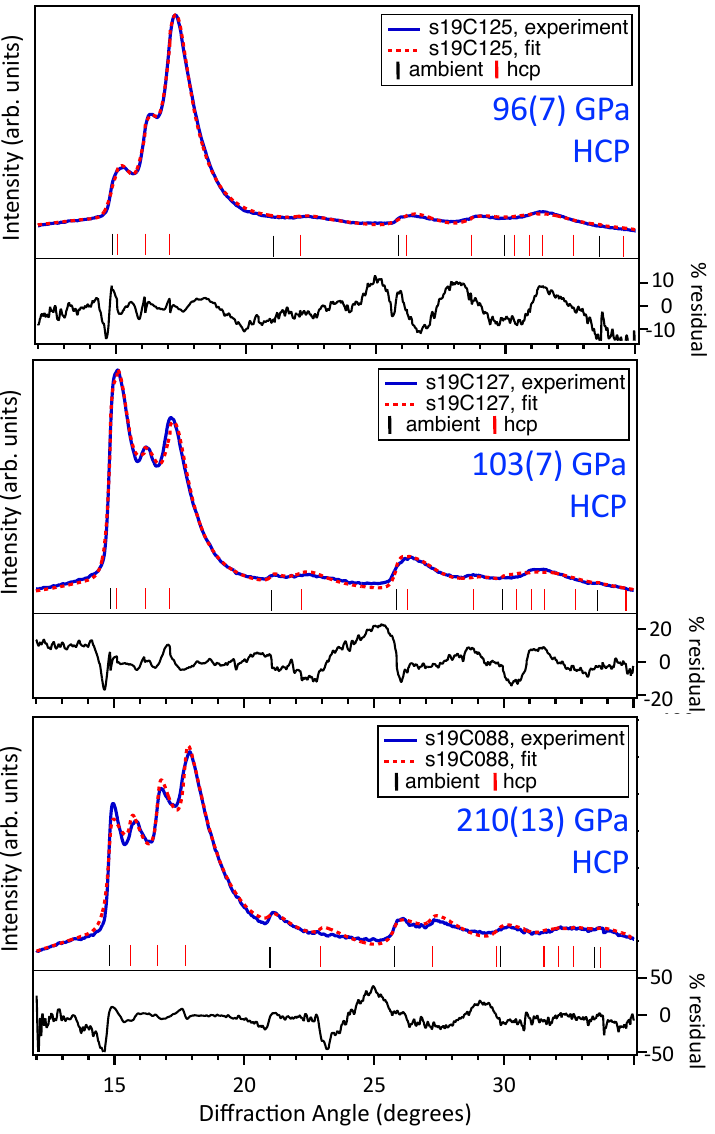}
\caption{\label{fig:Lineout_3} Azimuthally-integrated  profiles for experiments s19C125, s19C127, and s19C088 (blue) and fit using the HEXRD software (red dotted) \protect\cite{HEXRDgithubcp}.~All profiles include a contribution from the unshocked-volume (ambient fcc-phase) ahead of the shock front (black ticks), and the shock-compressed fcc-phase (red ticks) and/or bcc-phase (green ticks).~The \% difference between the experimental profile and the fit is shown on the right hand axis.}
\end{center}
\end{figure}

\noindent
Using XRD techniques our ability to constrain the extent of solid-liquid coexistence is limited by the ability to determine small volumes of liquid (melt onset) and solid (melt completion). For comparable volumes, scattering off liquid is more difficult to detect than a solid. Liquid diffraction is texture free (more powder like) and is broader in 2$\theta$ and lower in amplitude than scattering off a comparable volume of solid. In a mixed phase, if the signal from the liquid is below the threshold of detection then the analysis will show only compressed hcp as being present. In our transmission geometry shock compression in the mixed phase will result in a volume-integrated XRD pattern comprising of ambient-pressure bcc (ahead of the shock front), and compressed liquid+hcp behind (behind the shock front). Azimthual averaging of the XRD pattern (see Figs. \ref{fig:Lineout_1}, \ref{fig:Lineout_2}, and \ref{fig:Lineout_3}) results in an overlap of the three contributions. Determining small volumes of liquid is further complicated by the pink beam at DCS which results in overlapping of signal from closely spaced lines.
 
A recent study on Ge on the same experimental platform, using the same X-ray spectral distribution at the Dynamic Compression Sector has reported a detectable liquid volume thickness as low as $\sim$2 $\mu$m (Exp. 3 in Ref. \cite{Renganathan2023}). Since Fe is close in atomic number to Ge, we expect the detectability limit in Fe to be similar to the value in Ge. The thickness of the Fe samples used in Ref. \cite{Turneaure2020}
were 15µm, with a reported 89\% (or 13.35 $\mu$m) of compressed material. Based on the Ge study in Ref. \cite{Renganathan2023} it would be expected that at least 2 $\mu$m of Fe liquid thickness would be detectable (or 15\% of the compressed volume).

	\begin{figure}[h!]
	\begin{center}
	\includegraphics[width=0.8\columnwidth]{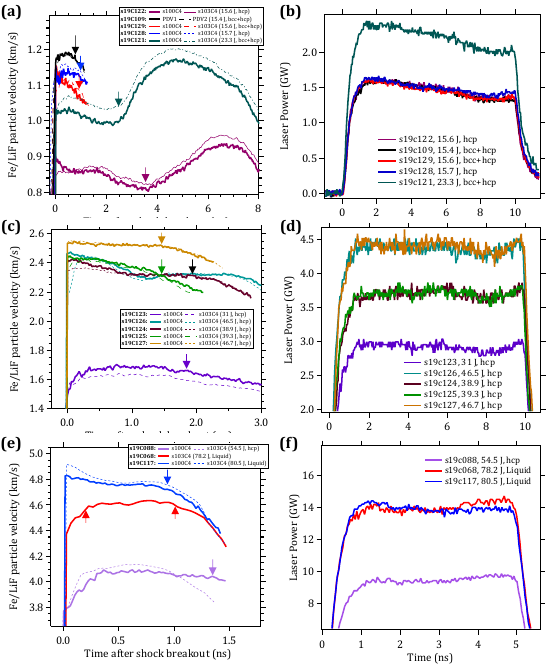}
	\caption{\textbf{Summary of Fe/LiF particle velocity traces} \textbf{(a), (c), (e)}, and the associated laser power versus time profiles \textbf{(b), (d), (f)}. The velocity traces were offset in time such that time=0 equates to the shock arrival time at the Fe/LiF interface. For some shots the point VISAR systems did not record usable data and instead a PDV VISAR record was used \protect\cite{dolan2006cp}. The shot number, laser energy for each shot, as well as the measured phases are listed in the legend. For pressure determination through impedance matching (as described in the main text) we considered the velocity distribution between the arrows (as indicated). For traces with only one arrow we accounted for velocities from t=0. When multiple velocity traces were obtained for a given shot we performed a weighted average and velocity distribution analysis over all the velocity information available (from multiple velocity records).
	}
	\label{fig:VISAR}
	\end{center}
	\squeezeup
	\squeezeup
\end{figure}

\begin{figure*}[h!]
\begin{center}
\includegraphics[width=0.7\textwidth]{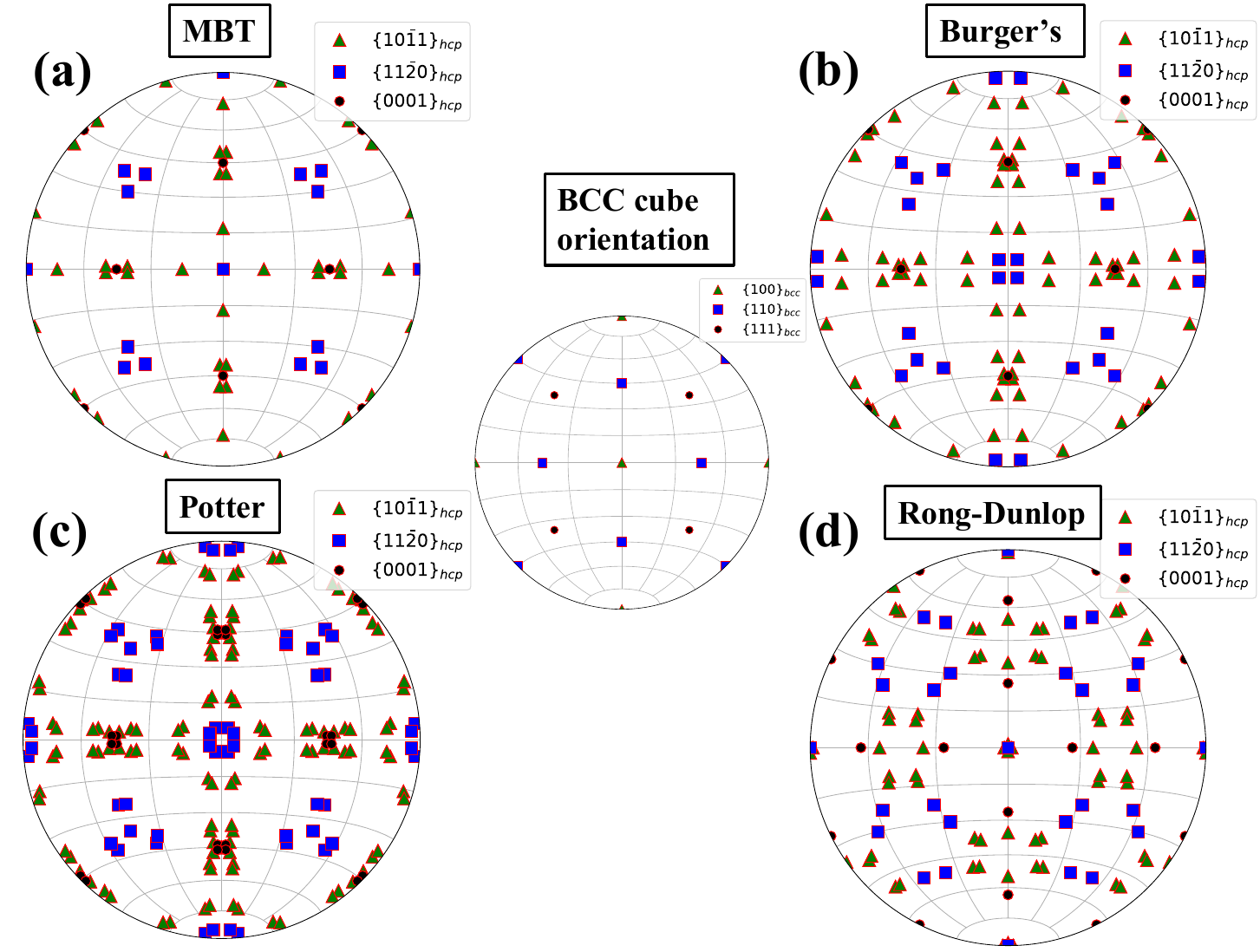}
\caption{\label{fig:pf_ors} \textbf{Pole figures for different mechanisms} Pole figures for some low index planes ambient cubic BCC phase in the cube orientation and high pressure hexagonal closed packed phase as a result of the different phase transition mechanisms: (a) Mao-Bassett-Takahashi OR given by the $(110)_{\text{bcc}}~\vert\vert~(0001)_{\text{hcp}}$ and $[001]_{\text{bcc}}~\vert\vert~[2\bar{1}\bar{1}0]_{\text{hcp}}$ (6 variants), (b) Burger's OR given by $(110)_{\text{bcc}}~\vert\vert~(0001)_{\text{hcp}}$ and $[1\bar{1}1]_{\text{bcc}}~\vert\vert~[11\bar{2}0]_{\text{hcp}}$ (12 variants), (c) Potter OR given by $(110)_{\text{bcc}}~\vert\vert~(1\bar{1}01)_{\text{hcp}}$ and $[1\bar{1}1]_{\text{bcc}}~\vert\vert~[11\bar{2}0]_{\text{hcp}}$ (24 variants) and (d) Rong-Dunlop OR given by $(021)_{\text{bcc}}~\vert\vert~(0001)_{\text{hcp}}$ and $[100]_{\text{bcc}}~\vert\vert~[11\bar{2}0]_{\text{hcp}}$ (12 variants).}
\end{center}
\end{figure*}

\begin{figure}[h!]
\centering
\includegraphics[width=0.75\textwidth]{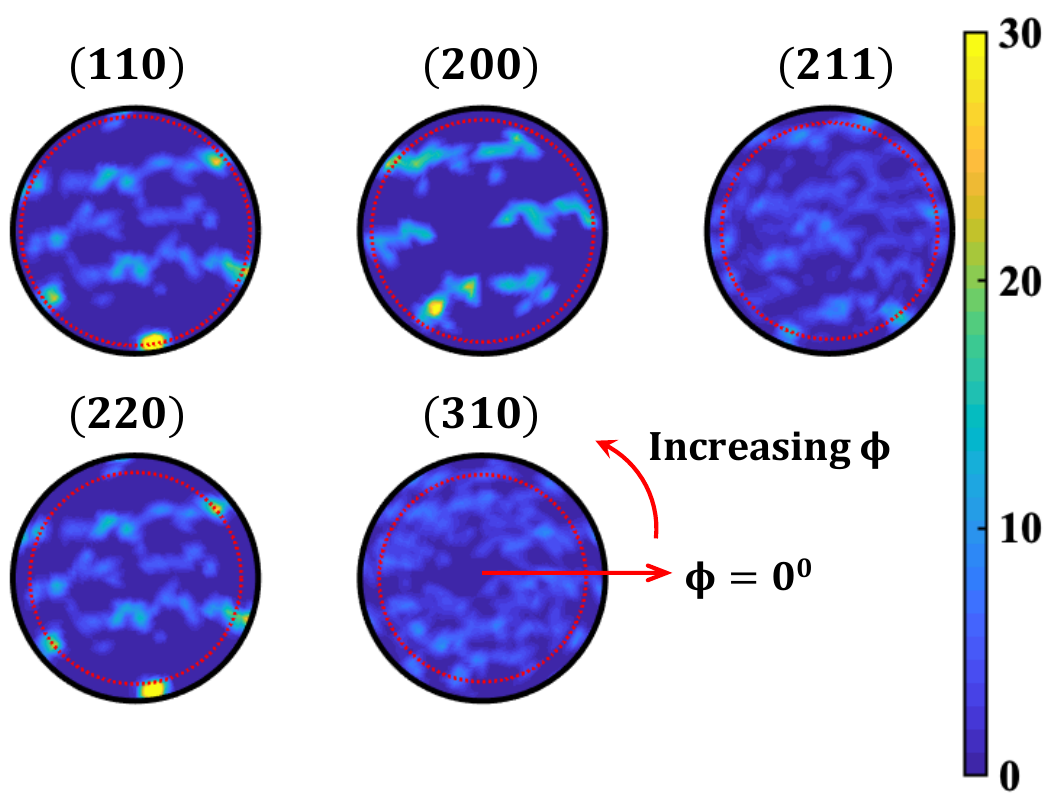}
\caption{\label{fig:bcc_calc_pfs} \textbf{Calculated pole figure for ambient \boldmath${\alpha-}$Fe} Calculated pole figures for the initial texture of the ambient cubic Fe sample. The red circle shows the region of the $2\theta-\phi$ space which are measured with our diffraction geometry. Color bar units are m.r.d. (multiples of random distribution), where 1 is random texture and $>$1 is non random. }
\end{figure}

\begin{figure}[h!]
\begin{center}
\includegraphics[width=0.75\textwidth]{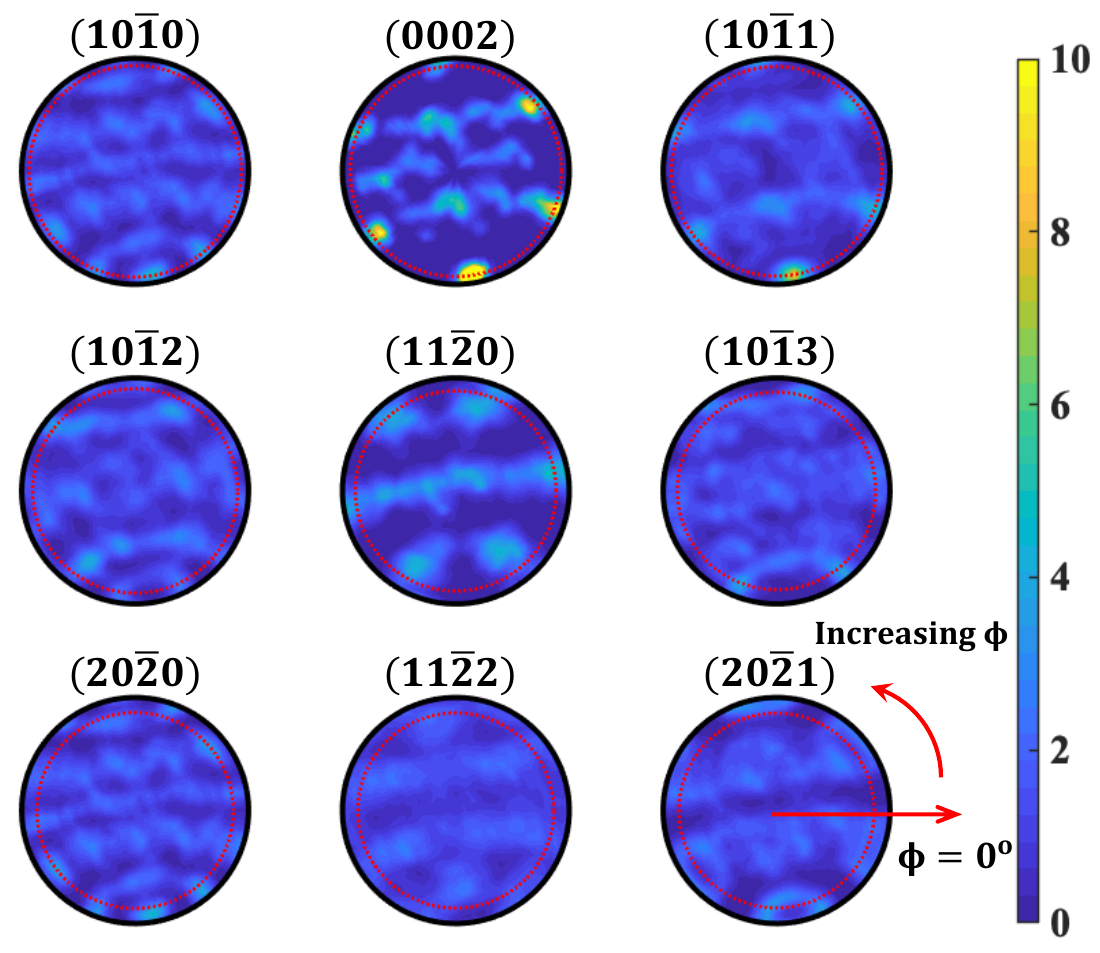}
\caption{\label{fig:bor_calc_pfs} \textbf{Calculated pole figure for Burger's OR for high-pressure \boldmath${\epsilon-}$Fe} Pole figures from applying the Burger's orientation relationship to the ambient sample texture shown in Fig.~\ref{fig:bcc_calc_pfs}. The red circle shows the part of the $2\theta-\phi$ space probed in our diffraction measurements. Color bar units are m.r.d. (multiples of random distribution), where 1 is random texture and $>$1 is non random.}
\end{center}
\end{figure}

\begin{figure}[h!]
\begin{center}
\includegraphics[width=0.75\textwidth]{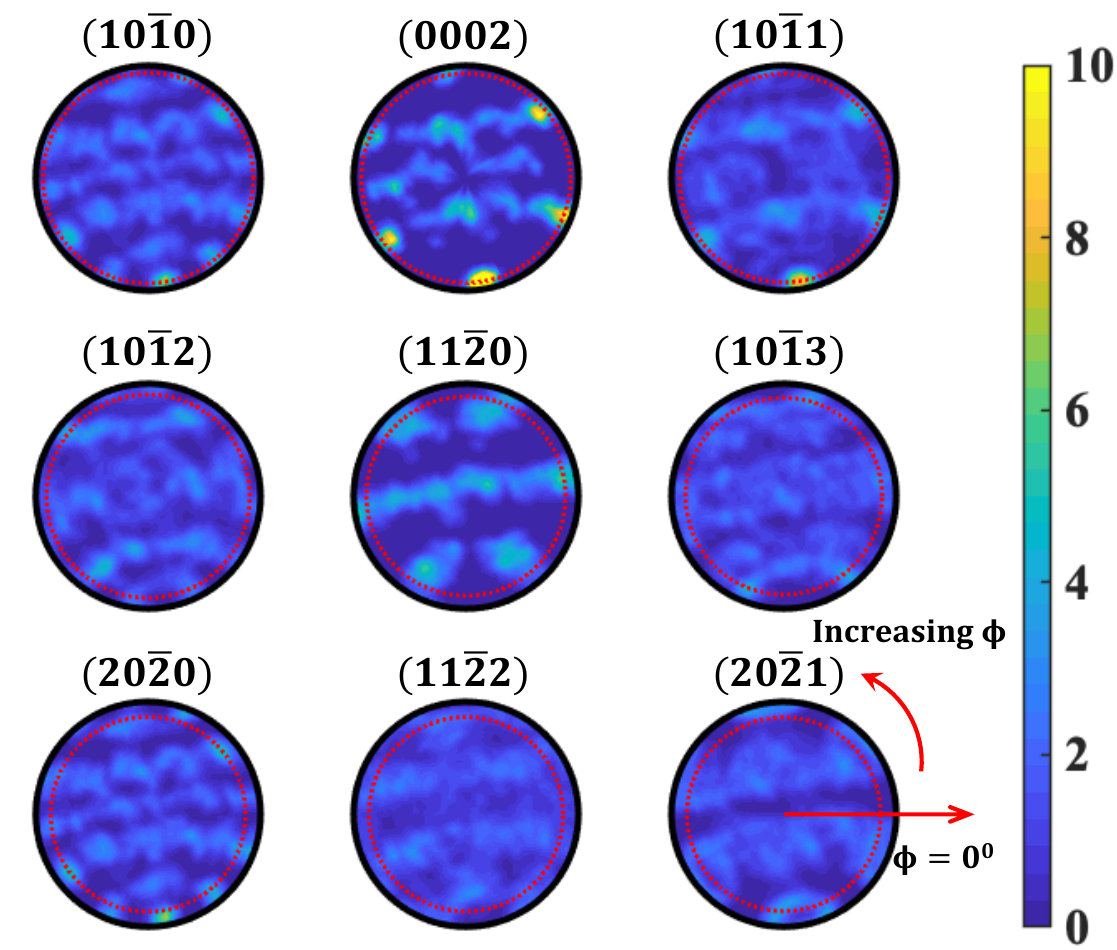}
\caption{\label{fig:mbt_calc_pfs} \textbf{Calculated pole figure for MBT OR for high-pressure \boldmath${\epsilon-}$Fe} Pole figures from applying the Mao-Bassett-Takahashi orientation relationship to the ambient sample texture shown in Fig.~\ref{fig:bcc_calc_pfs}. The red circle shows the part of the $2\theta-\phi$ space probed in our diffraction measurements. Color bar units are m.r.d. (multiples of random distribution), where 1 is random texture and $>$1 is non random.}
\end{center}
\end{figure}

\begin{figure}[h!]
\begin{center}
\includegraphics[width=0.75\textwidth]{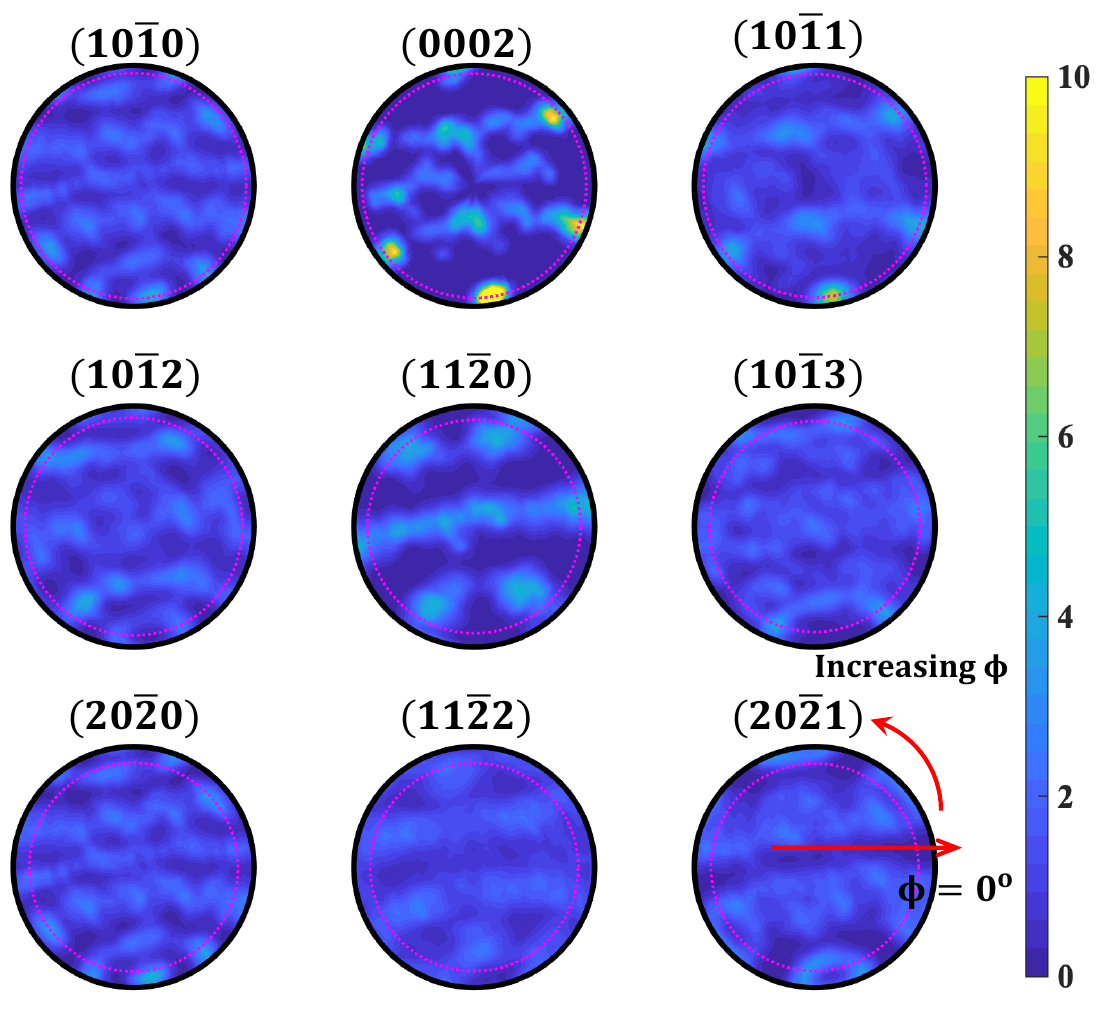}
\caption{\label{fig:potter_calc_pfs} \textbf{Calculated pole figure for Potter's OR for high-pressure \boldmath${\epsilon-}$Fe} Pole figures from applying the Potter orientation relationship to the ambient sample texture shown in Fig.~\ref{fig:bcc_calc_pfs}. The red circle shows the part of the $2\theta-\phi$ space probed in our diffraction measurements. Color bar units are m.r.d. (multiples of random distribution), where 1 is random texture and $>$1 is non random.}
\end{center}
\end{figure}

\begin{figure}[h!]
\begin{center}
\includegraphics[width=0.75\textwidth]{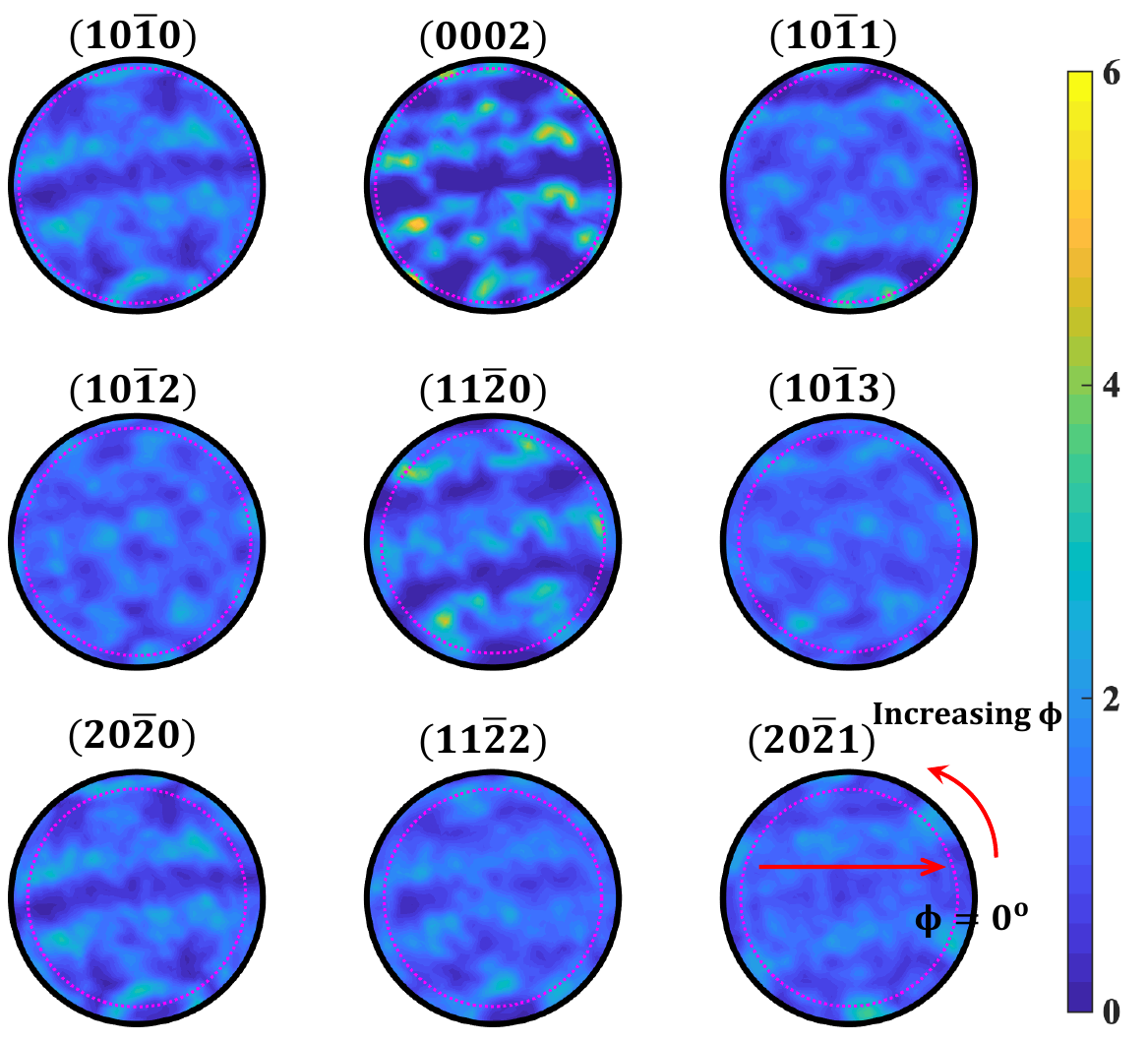}
\caption{\label{fig:rong_calc_pfs} \textbf{Calculated pole figure for Rong-Dunlop OR for high-pressure \boldmath${\epsilon-}$Fe} Pole figures from applying the Rong-Dunlop orientation relationship to ambient the sample texture shown in Fig.~\ref{fig:bcc_calc_pfs}. The red circle shows the part of the $2\theta-\phi$ space probed in our diffraction measurements. Color bar units are m.r.d. (multiples of random distribution), where 1 is random texture and $>$1 is non random.}
\end{center}
\end{figure}

\begin{figure}
\begin{center}
\includegraphics[width=\textwidth]{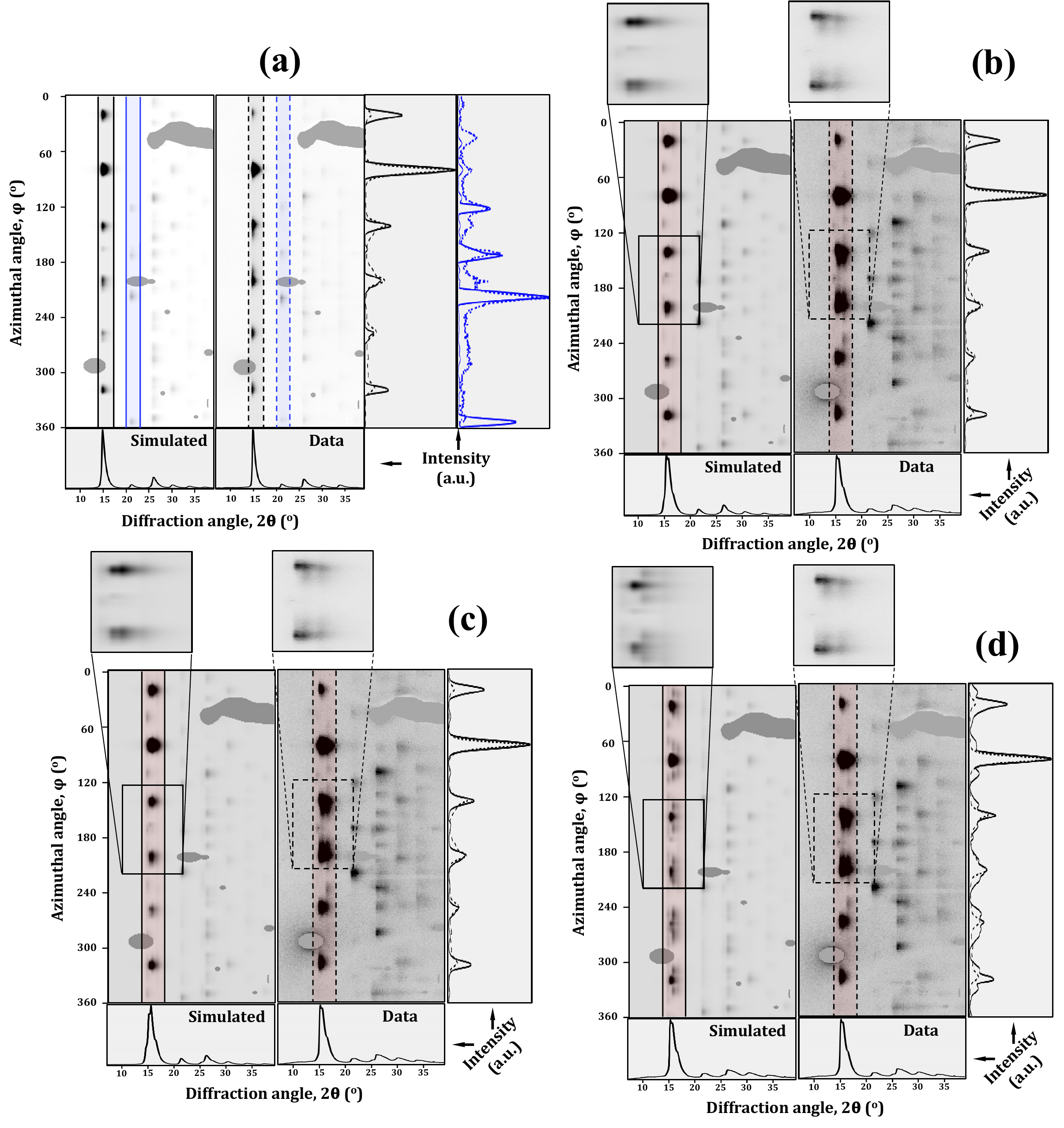}
\caption{\textbf{Simulated diffraction patterns} Comparison of simulated diffraction signal with the experimental diffraction signal for (a) ambient $\alpha-$Fe and high-pressure hcp phase following (b) Mao-Bassett-Takahashi (MBT) (c) Potter and (d) Rong-Dunlop ORs. The azimuthal lineout for the band between $14^{\circ}-18^{\circ}$ diffraction angle is shown on the right for the simulated (solid) and experimental (dashed) signal. The azimuthally averaged lineout is shown at the bottom. The inset show presents a high-resolution view of a small section of the simulated and experimental data. We are not able to distinguish between the Burger's, MBT and Potter's OR based on our data. However, we are able to rule out the Rong-Dunlop OR. }
\label{fig:texture_other}
\end{center}
\end{figure}

\clearpage
\bibliographystyle{elsarticle-num}
\bibliography{references}

\end{document}